\documentclass[10pt,journal,compsoc]{IEEEtran}
\usepackage{amsmath,amsfonts}
\usepackage{algorithmic}
\usepackage{algorithm}
\usepackage{array}
\usepackage{textcomp}
\usepackage{stfloats}
\usepackage{url}
\usepackage{verbatim}
\usepackage{graphicx}
\usepackage{lineno}
\usepackage{multirow}
\usepackage{float}
\usepackage{graphicx}
\usepackage{amsmath}
\usepackage{color}
\usepackage{soul}
\usepackage{booktabs}
\usepackage{threeparttable}
\usepackage{caption}
\usepackage{indentfirst}
\usepackage{color}
\usepackage{bm}
\usepackage[cmyk]{xcolor}
\usepackage[T1]{fontenc}
\usepackage{orcidlink}
\ifCLASSOPTIONcompsoc
\usepackage[nocompress]{cite}
\else
\usepackage{cite}
\fi
\ifCLASSINFOpdf
\else
\fi
	\hypersetup{
		colorlinks=true,
		linkcolor=black,
		citecolor=black,
		hidelinks
	}
	
\begin{document}
		%
		\title{IOPS: An Unified SpMM Accelerator Based on Inner-Outer-Hybrid Product}
		%
		%
		%
		%
		
		\author{Wenhao~Sun~\orcidlink{0009-0003-6118-8174},
			Wendi~Sun~\orcidlink{0009-0002-3198-2345},
			Song~Chen,~\IEEEmembership{Member,~IEEE,}
			and~Yi~Kang
			
			\IEEEcompsocitemizethanks{
				\IEEEcompsocthanksitem This work was supported in part by the Strategic Priority Research Program of Chinese Academy of Sciences, Grant No. XDB44000000 ,in part by CAS Project for Young Scientists in Basic Research under grant No. YSBR-029k, and in part by the National Key R\&D Program of China under grant No. 2019YFB2204800.
				The authors want to thank Information Science Laboratory Center of USTC for the hardware \& software services.\
				\IEEEcompsocthanksitem Wenhao Sun and Wendi Sun are affiliated with the School of Microelectronics, University of Science and Technology of China, Hefei,
				Anhui, 30332 China.\
				\IEEEcompsocthanksitem Song Chen and Yi Kang are affiliated with the School of Microelectronics,  University of Science and Technology of China, and also with the Institute of Artificial Intelligence, Hefei Comprehensive National Science Center,  Hefei,
				Anhui, 30332 China. \protect\\(Corresponding author: Song chen, E-mail: songch@ustc.edu.cn) \
				
			}
			
			\thanks{ }}

	\IEEEtitleabstractindextext{%
		\begin{abstract}
			Sparse matrix multiplication (SpMM) is widely applied to numerous domains, such as graph processing, machine learning, and data analytics. However, inner product based SpMM induces redundant zero-element computing for mismatched nonzero operands, while outer product based approach lacks input reuse across Process Elements (PEs) and poor output locality for accumulating partial sum (psum) matrices. Besides, current works only focus on sparse-sparse matrix multiplication (SSMM) or sparse-dense matrix multiplication (SDMM), rarely performing efficiently for both. To address these problems, this paper proposes an unified SpMM accelerator, called IOPS, hybridizing inner with outer products. It reuses the input matrix among PEs with inner product dataflow, and removes zero-element calculations with outer product approach in each PE, which can efficiently process SSMM and SDMM. Moreover, an address mapping method is designed to accumulate the irregular sparse psum matrices, reducing the latency and DRAM access of psum accumulating. Furthermore, an adaptive partition strategy is proposed to tile the input matrices based on their sparsity ratios, effectively utilizing the storage of architecture and reducing DRAM access. Compared with the SSMM accelerator, SpArch, we achieve $1.7\times$\textasciitilde$6.3\times$ energy efficiency and $1.2\times$\textasciitilde$4.4\times$ resource efficiency, with $1.4\times$\textasciitilde$2.1\times$ DRAM access saving.
		\end{abstract}
		
		\begin{IEEEkeywords}
			
			Sparse matrix multiplication, accelerator, Inner-Outer Product.
		\end{IEEEkeywords}}

	\maketitle

	\IEEEdisplaynontitleabstractindextext

	%
	\IEEEpeerreviewmaketitle

	\IEEEraisesectionheading{\section{Introduction}\label{sec:introduction}}
	SpMM is the key computing kernel for many applications, such as shortest-paths search\cite{10.5555/3118786.3119254}, recommendation systems \cite{ 10.1145/3219819.3219890}, molecular dynamics\cite{10.5555/866839}, and graph computing \cite{kipf2017semisupervised}. Since the density of sparse matrices is as low as $10^{-3}$\textasciitilde$10^{-6}$ \cite{snapnets}, removing zero-element calculations can greatly enhance system performance and energy efficiency.\\
	\indent\setlength{\parindent}{1em}Currently, SpMM is calculated based on inner or outer product approach. For inner product, its computation pattern is dot product between each row of matrix A and each column of matrix B to obtain an element of output matrix, which can fully reuse the output matrix. To increase the reuse of input matrices, inner product calculations are often mapped to systolic arrays, such as TPU \cite{ 8192463}, RASA \cite{ 9586257}. However, the mismatched positions of nonzero elements in the input matrix introduce a large number of zero element calculations, causing degradation of performance and energy efficiency. Zhu et al\cite{94} and RP-GCN \cite{ 9774537} adopt an offline clustering approach to compress the sparse adjacency matrix into multiple dense matrices. However, this preprocessing step takes excessive time in the whole process, resulting in low performance of overall system. HyGCN\cite{9065592} proposes a static graph partition for data reuse and dynamic sparsity elimination to reduce redundant accesses. SPOTS \cite{ 10.1145/3532863} removes the all-zero rows or columns of input matrix through bitmap mask. But there still remain many zeros in the processed matrix, causing inefficient performance improvement. \\
	\indent\setlength{\parindent}{1em}For the outer product approach, the computation pattern is outer product between the columns of matrix A and the rows of matrix B to generate a psum matrix. This approach can overcome the irregular distribution of nonzero elements and theoretically skip all calculations of zero elements, thus significantly improving performance. However, the distribution of nonzero elements in each psum matrix is irregular, which involves enormous time and energy costs to accelerators, such as Dalton et al \cite{49}, Liu et al \cite{51}, Lee et al \cite{lee}, MOSCON \cite{10089919}, and OuterSPACE \cite{8327050}. To address the irregularity of psum matrix, SpArch \cite{9065428} condenses the psum matrix with a Huffman Tree scheduler and accumulates them with a merge tree, accelerating the process of psum accumulation. AWB-GCN\cite{9252000} and GCNAX\cite{9407104} apply row remapping strategy to accumulate psum more regularly. MatRaptor \cite{ 9251978} proposes a row-wise product approach, which computes the output matrix by row and achieves high resource efficiency. However, These works neglect data reuse among PEs, resulting in longer delay of data loading and lower overall performance. \\
	\indent\setlength{\parindent}{1em}The applications based on SpMM include not only SSMM, such as interior point methods\cite{10.1145/1137856.1137866}, searching algorithms \cite{10.1109/MCSE.2008.45}, and matching algorithms \cite{RABIN1989557}, but also SDMM, such as  GCNs\cite{kipf2017semisupervised}, GraphSage\cite{10.5555/3294771.3294869}, GINConv\cite{Xu2018HowPA}, and GAT\cite{GAT}. However, some work only accelerate the former, such as Zhu et al\cite{94}, AWB-GCN\cite{9252000}, and GCNAX\cite{9407104}, which can only exploit the sparsity of one matrix. Other works only focus on the acceleration of SSMM, such as OuterSPACE \cite{8327050}, SpArch \cite{9065428}, and  MatRaptor \cite{ 9251978}, which aggravates the latency of input loading and psum accumulating due to data increment in SDMM. HyGCN\cite{9065592} supports SSMM and dense-dense matrix multiplication (DDMM) by integrating two different computing engines, causing extra resource consumption for the interface and imbalanced workload  in some cases between two engines.  \\
	\indent\setlength{\parindent}{1em}To address these problems above, this paper proposed an unified SpMM accelerator based on inner-outer-hybrid product (IOHP), adaptive for various SpMM processing. Our main contributions are as follow: \\ 
	\indent\setlength{\parindent}{1em}1). To address the mismatched operands of inner product and lack of input reuse of outer product, we proposed an IOHP method for SpMM, which reuses the input matrix across PEs with inner product and removes zero-element calculations with outer product in each PE, and collaborated with an address mapping method to accumulate the irregular psum matrices, reducing the latency and DRAM access of psum accumulating.  \\
	\indent\setlength{\parindent}{1em}2). To support both SSMM and SDMM computing, we designed an unified SpMM accelerator, called IOPS, whose procedure for SSMM consists of three pipeline stages: encoding input matrices with RP-CSC (Row Partition CSC) and CP-CSR (Column Partition CSR) format, calculating psum based on IOHP, accumulating psum through address mapping. For SDMM, we omit the last stage and reorganize the psum buffers to reduce DRAM access.   \\
	\indent\setlength{\parindent}{1em}3). To optimize the DRAM access, we proposed an adaptive partition strategy to tile the input matrices based on their sparsity and size, effectively utilizing the storage of architecture and reducing DRAM access. \\ 
	\indent\setlength{\parindent}{1em}Compared with inner product based accelerator, SPOTS\cite{10.1145/3532863}, we achieve $14.3\times$\textasciitilde$43.7\times$ performance and $7.2\times$\textasciitilde$21.9\times$ energy efficiency due to the IOHP approach. Compared with outer product based accelerator for SDMM, GCNAX\cite{9407104}, we achieve a similar performance in the SDMM benchmark and speed up by $9.8\times$\textasciitilde$12.1\times$ in the SSMM benchmark because of the unified computing mode. For the outer product based accelerator for SSMM, SpArch\cite{9065428}, we achieve $1.3\times$\textasciitilde$4.8\times$ performance and $1.7\times$\textasciitilde$6.3\times$ energy efficiency for the efficient reuse of input matrices.
	
	\section{Prelimery}
	
	\subsection{Compression Format of Sparse Matrix}
	
\begin{figure}[H]
		\centering
		\includegraphics[width=0.99\linewidth]{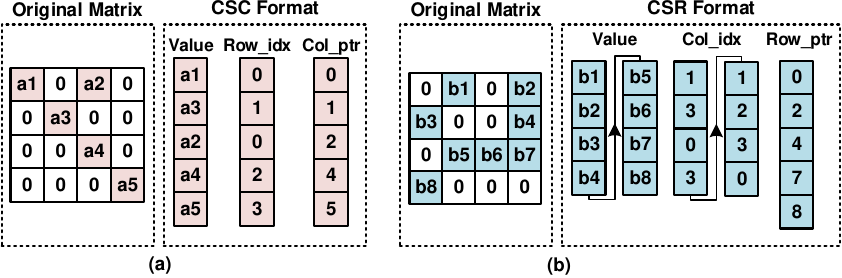}
		\caption{Compression format of sparse matrix. (a)CSC format. (b)CSR format.}
		\label{fig1}
	\end{figure}

	Generally, the density of sparse matrix is quiet low. To reduce the storage cost, it is usually compressed with Compressed-Sparse-Column (CSC) format or Compressed-Sparse-Row (CSR) format, as shown in Fig.\ref{fig1}. 
	For the matrix of CSC format, it consists of $value$, $row\_idx$, and $col\_ptr$, where $row\_idx$ records the row index of element and $col\_ptr$ represents the start point of element in each column. For the matrix of CSR format, it consists of $value$, $col\_idx$, and $row\_ptr$, where $col\_idx$ records the column index of each element and $col\_ptr$ represents the start point of element in each row. Due to the lower storage cost, the matrices of CSC and CSR format are the appropriate input for SpMM.

%
	\begin{figure*}
		\centering
		\includegraphics[width=0.99\linewidth]{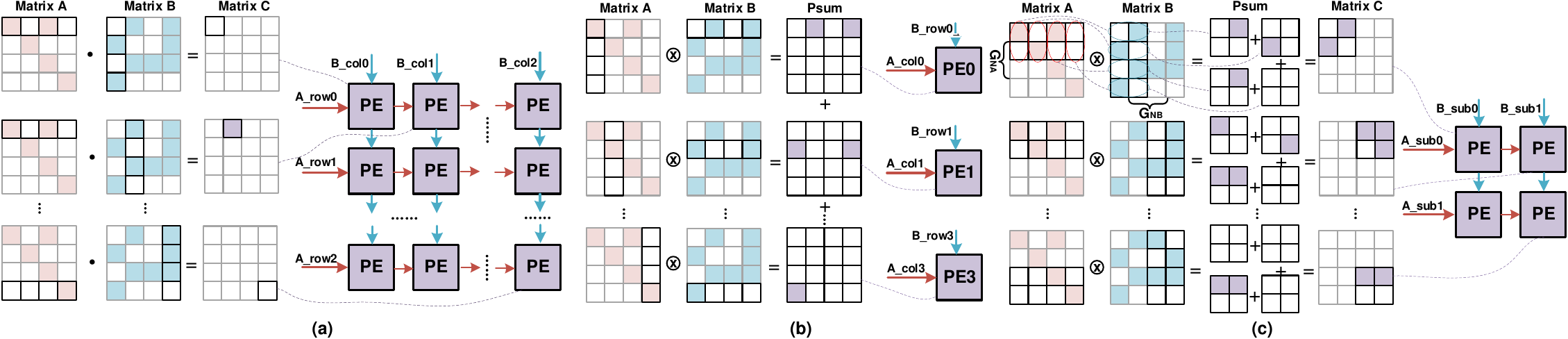}
		\caption{Computing approach of SpMM. (a)Inner product. (b)Outer product. (c)IOHP.}
		\label{fig2}
	\end{figure*}
	\subsection{Inner Product}

	Inner product is widely applied for regular matrix multiplication, which conducts a dot product between one row of matrix A and one column of matrix B, as shown in Fig.\ref{fig2}(a). 
	Assuming that the dimension of input matrix A, B, and output matrix C are $M\times K$, $K\times N$, $M\times N$, respectively, this approach can be mapped into a systolic array with high data reuse rate, where each row of matrix A, $A\_row_i (i=0,1,..,M-1)$, is shared by one row of PEs, and each column of matrix B, $B\_col_j (j=0,1,..,N-1)$, is shared by one column of PEs. For the PE of $i$-$th$ row and $j$-$th$ column, the operation can be expressed by Equ.\ref{equ2}, obtaining one final result of matrix C. \\
	\indent\setlength{\parindent}{1em}Although the inner product can reuse the input matrices efficiently, the mismatched indexes of matrix A and B in the PE array will induce a large amount of zero-element computing for SpMM, degrading the performance dramatically. Thus, how to accelerate the inner product approach is worth exploring.
	\begin{equation}
		\label{equ2}
		\begin{aligned}
			C[i,j] = \sum\limits_{x=1}^{K-1}A[i,x]*B[x,j]
		\end{aligned}
	\end{equation}

	\subsection{Outer Product}

	Outer product can naturally avoid the mismatched indexes, where an outer product is performed between one column of matrix A and one row of matrix B, as shown in Fig.\ref{fig2}(b). In this approach, the $i$-$th (i=0,1,..,K-1)$ PE receives the $i$-$th$ column of matrix A, $A\_col_i$, and $i$-$th$ row of matrix B, $B\_row_i$, to obtain one psum matrix, as expressed by Euq.\ref{equ3}. For each PE, it only needs to compute the non-zero elements, which achieves higher performance compared with inner product.\\
	\indent\setlength{\parindent}{1em}Although outer product can accelerate SpMM efficiently, there is no input matrix reusing among PEs, which requires $N\times$ bandwidth of loading input compared with inner product for an $N\times N$ PE array. Once the data loading speed is slower than the PE computation, the overall performance will also be degraded. Thus, how to enhance the input reusing of outer product is also a key challenge.  
	\begin{equation}
		\label{equ3}
		\begin{aligned}
			psum[a,b] = A[a,i]*B[i,b]\  (a\in[0,M),b\in[0,N))
		\end{aligned}
	\end{equation}   
	
	\section{Inner-Outer-Hybrid Product}
	
	\subsection{Computing flow for SSMM}
\begin{figure}[H]
		\centering
		\includegraphics[width=0.99\linewidth]{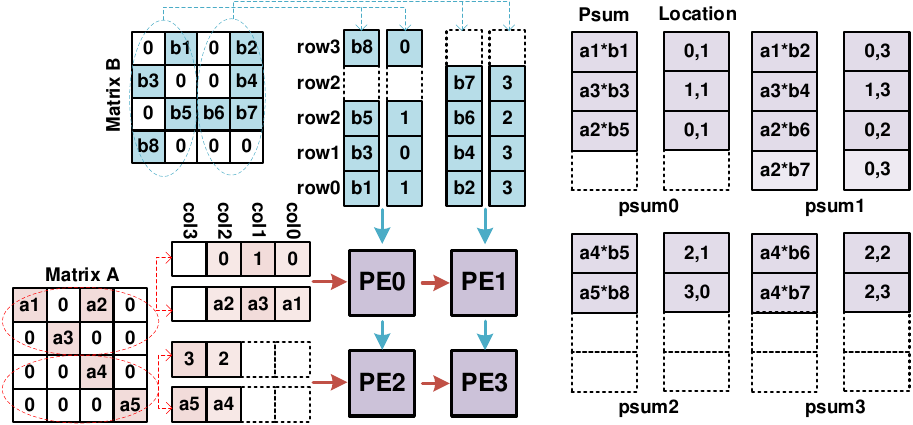}
		\caption{An example of SSMM based on IOHP.}
		\label{fig3}
	\end{figure}

	 To solve the problem of inner and outer product, we proposed an IOHP approach for SpMM acceleration, as shown in Fig.\ref{fig2}(c). Firstly, the matrices A and B are respectively divided to $G_{NA}$ groups in the row dimension and $G_{NB}$ groups in the column dimension, and the specific partition method is explained in Section.5. Assuming that the size of tiled matrix $A\_sub_i (i=0,1,..,G_{NA}-1)$ and $B\_sub_j (j=0,1,..,G_{NB}-1)$ is $M_t\times K$ and $K\times N_t$ respectively, each $A\_sub_i$ and $B\_sub_j$ can be shared by the PEs in the same row and column respectively  as the inner product approach. Then, the PE in $i$-$th$ row and $j$-$th$ column performs an outer product between a $A\_sub_i$ and $B\_sub_j$, which is expressed in Euq.\ref{equ4} and can skip most zero-element calculations as outer product approach. Finally, we can both reuse the input matrix and improve the performance efficiently based on IOHP. 
	 \begin{equation}
	 	\label{equ4}
	 	\begin{aligned}
	 		C[a,b] = \sum\limits_{x=1}^{K-1}A[a,x]*B[x,b]\ \ \ \ \ \ \ \ \ \ \ \ \ \ \ \ \  \\
	 		(a\in[i*M_t,(i+1)*M_t),b\in[j*N_t,(j+1)*N_t)) 
	 	\end{aligned}
	 \end{equation}  
	 \indent\setlength{\parindent}{1em}An example of IOHP computing is shown in Fig.\ref{fig3}. First, the $4\times4$ input matrix A and B are divided into two $2\times4$ and $4\times2$ submatrices. Then, we extract the non-zero elements and their corresponding location from these submatrices. Finally, we load these input elements to perform outer product in each PE and obtain all psum elements. Compared with inner product, we achieve $1.8\times$ speedup. Compared with outer product, we save $2\times$ bandwidth of loading input matrix. 
	
	\subsubsection{Encoding Format}
\begin{figure}[H]
		\centering
		\includegraphics[width=0.8\linewidth]{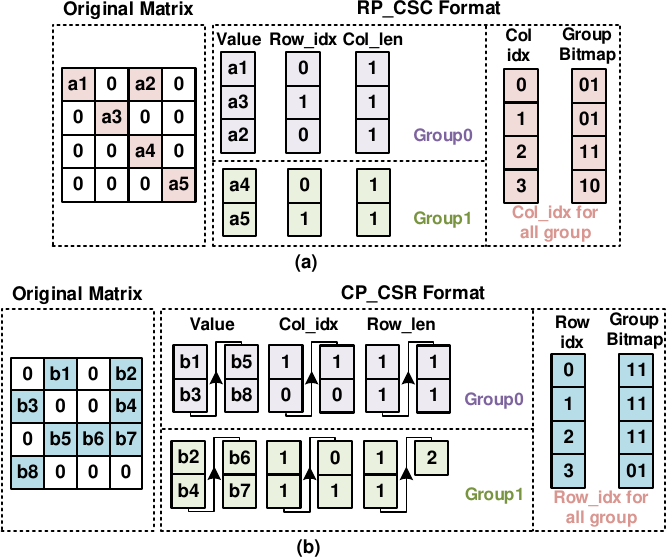}
		\caption{Encoding format of matrix A and B. (a)RP-CSC format of matrix A. (b)CP-CSR format of matrix B.}
		\label{fig4}
	\end{figure}

	To fit with the computing flow of IOHP, we need to transfer the original input matrices of CSC and CSR format to RP-CSC and CP-CSR respectively, as shown in Fig.\ref{fig4}. For matrix A, apart from the value and row index, we need to record the length of non-zero elements in each column. Besides, after row partition, not every group of submatrix A contains non-zero elements in each column. Thus, we apply a group bitmap to record the empty/non-empty state of each group in the current column. Finally, submatrix A of RP-CSC format consists of $value$, $row\_idx$, $col\_len$,  $col\_idx$, and $group\_bitmap$, where the former three are stored in each group and the latter two are stored once for all groups. Similarly, submatrix B of CP-CSR format consists of $value$, $col\_idx$, $row\_len$,  $row\_idx$, and $group\_bitmap$.
	\indent\setlength{\parindent}{1em}The specific flow of encoding is shown in Tab.\ref{tab1}. Since the sizes of input matrices in the real world are unpredictable, we need to tile them further in the row and column dimension. The detail is discussed in Section.5. After partition, we obtain $T_M\times T_K$ blocks of $G_{NA}$ groups of $M_t\times K_t$ submatrix A, and $T_K\times N_K$ blocks of $G_{NB}$ groups of $K_t\times N_t$ submatrix B in the 1\textasciitilde3-th line. For $G_{NA}$ groups of $M_t\times K_t$ submatrix A, we encode them column-wise in the 5-th line. First, we locate the first element with $col\_ptr\_A$ in the 7-th line. If the $row\_idx\_A$ is in the range of row index of the current block, we figure out which group it belongs to in the 8-th line. Then, we copy the $value$ and $row\_idx$ to the corresponding group in the 9-th line. The $col\_len\_sA$ self-increases one and the $group\_bitmap\_sA$ is set to $1'b1$ in the 10\textasciitilde11-th line. When the $row\_idx\_A$ is out of range, we update the original $col\_ptr\_A$ with the current location and record the $col\_idx\_sA$ in the 12\textasciitilde13-th line. After that, we repeat the above operations until all columns are traversed. Finally, we record the length of non-empty columns for all groups, $col\_all\_len\_sA$ in the 16-th line. For $G_{NB}$ groups of $K_t\times N_t$ sub-matrix B, we encode them row-wise with similar operations in the 17\textasciitilde27-th line.
	\begin{table}[H]\centering
		\caption{Encoding flow of matrix A and B}
		\label{tab1}
		\begin{tabular}{lll}
			\toprule
			\textbf{Input}: Matrix A of CSC: $value\_A$, $row\_idx\_A$, $col\_ptr\_A$; \\
			Matrix B of CSR: $value\_B$, $col\_idx\_B$, $row\_ptr\_B$;\\
			\# of tiling matrix A/B: $T_M$, $T_K$, $T_N$; \\
			\# of group submatrix A/B: $G_{NA}$, $G_{NB}$; \\
			Size of submatrix A/B: $M_t\times K_t$, $K_t\times N_t$.\\
			\textbf{Output}: submatrix A of RP\_CSC: $value\_sA$, \\
			$row\_idx\_sA$, $col\_len\_sA$, $col\_idx\_sA$, $group\_bitmap\_sA$;\\
			submatrix B of CP\_CSR: $value\_sB$, $col\_idx\_sB$, \\
			$row\_len\_sB$, $row\_idx\_sB$, $group\_bitmap\_sB$.  \\
			\midrule
			\textbf{1}\hspace{0.3cm} $for (a=0; a<T_M; a=a+1)$ \\
			\textbf{2}\hspace{0.6cm} 	$for (b=0; b<T_M; b=b+1)$ \\
			\textbf{3}\hspace{0.9cm} 		$for(c=0;c<T_K; c=c+1)$ \\
			\textbf{4}\hspace{0.9cm}        $Reset\ all\ variable$ \\
			\textbf{5}\hspace{1.2cm} 			$for(d=0;d< K_t;d=d+1)$ \\
			\textbf{}\hspace{0.3cm}$///////\ \bm{RP-CSC\ Encoding\ of\ matrix\ A}\ //////$				\\
			\textbf{6}\hspace{1.5cm} 				$e = 0$\\
			\textbf{7}\hspace{1.5cm}				$while(row\_idx\_A\ in\ a-th\ row-block)$  \\
			\textbf{8}\hspace{1.8cm} 					$group\_id = row\_idx\_A/M_t$  \\
			\textbf{9}\hspace{1.8cm}					$value\_sA[group\_id] = value\_A$ \\
			\textbf{10}\hspace{1.7cm}					$row\_idx\_SA[group\_id] = row\_idx\_A$ \\
			\textbf{11}\hspace{1.7cm}					$col\_len\_sA[group\_id]++$ \\
			\textbf{12}\hspace{1.7cm}					$group\_bitmap\_sA[group\_id] = 1'b1$ \\
			\textbf{13}\hspace{1.7cm}					$e++$  \\
			\textbf{14}\hspace{1.4cm}				$col\_all\_id++$ \\
			\textbf{15}\hspace{1.4cm}				$col\_ptr\_A += e, col\_idx\_sA = d$ \\
			\textbf{16}\hspace{1.0cm}			$col\_all\_len\_sA = col\_all\_id$ \\
			\textbf{}\hspace{0.3cm}$//////\ \bm{CP-CSR\ Encoding\ of\ matrix\ B}\ //////$	\\
			\textbf{17}\hspace{1.4cm}				$f = 0$ \\
			\textbf{18}\hspace{1.4cm}				$while(col\_idx\_B\ in\ b-th\ column-block)$ \\
			\textbf{19}\hspace{1.7cm}					$group\_id = col\_idx\_B/N_t$ \\
			\textbf{20}\hspace{1.7cm}					$value\_sB[group\_id] = value\_B$ \\
			\textbf{21}\hspace{1.7cm}					$col\_idx\_sB[group\_id] = col\_idx\_B$ \\
			\textbf{22}\hspace{1.7cm}					$row\_len\_sB[group\_id]++$ \\
			\textbf{23}\hspace{1.7cm}					$group\_bitmap\_sB[group\_id] = 1'b1$ \\
			\textbf{24}\hspace{1.7cm}					$f++$ \\
			\textbf{25}\hspace{1.4cm}				$row\_all\_id++$ \\
			\textbf{26}\hspace{1.4cm}				$row\_ptr\_B += f$, $row\_idx\_SB = d $ \\
			\textbf{27}\hspace{1.0cm}			$row\_all\_len\_B = row\_all\_id$\\
			\bottomrule
		\end{tabular}
	\end{table}
	
	\subsubsection{Psum Calculation}
	
\begin{figure*}
		\centering
		\includegraphics[width=0.8\linewidth]{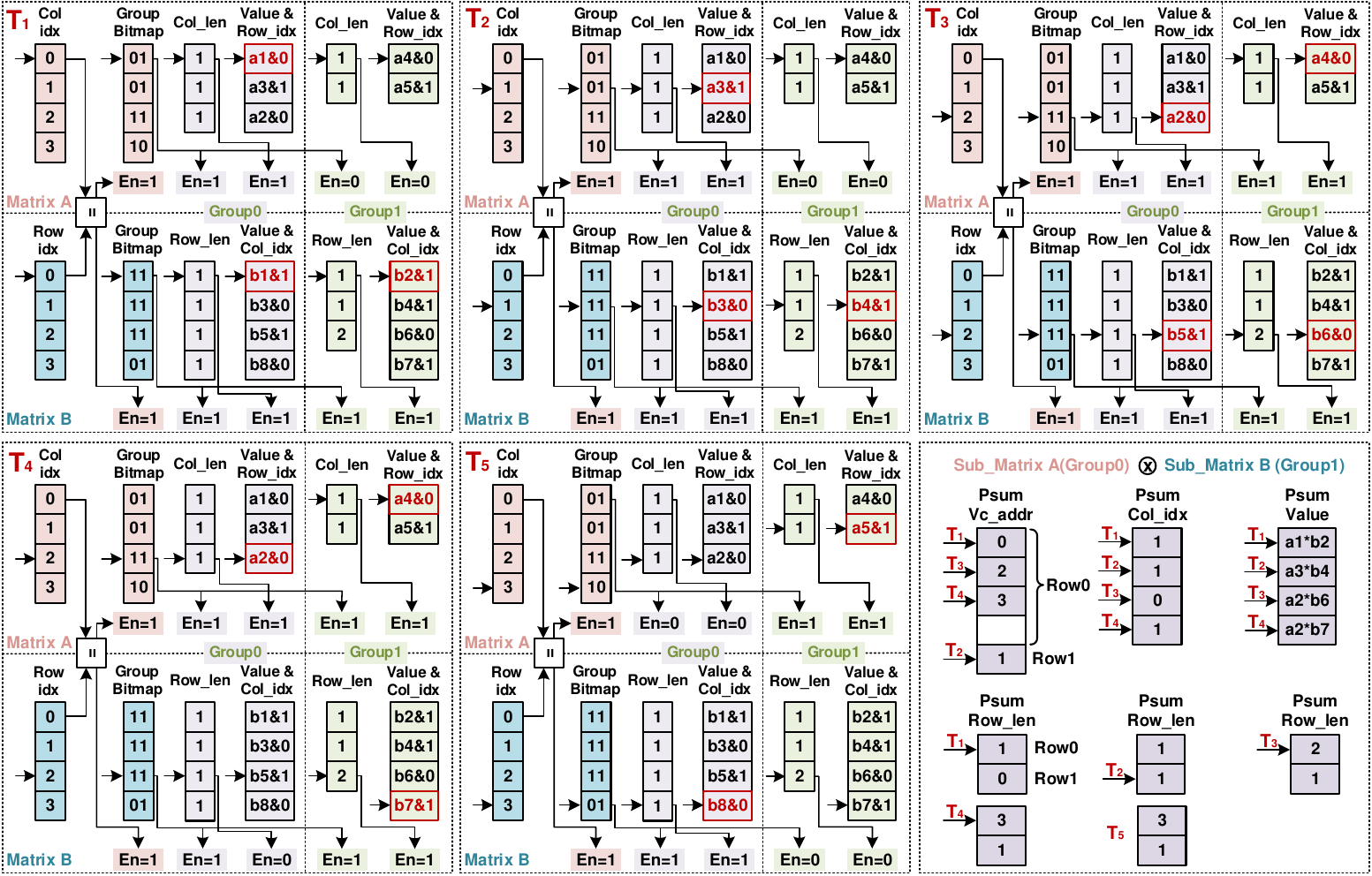}
		\caption{An example of psum calculation.}
		\label{fig5}
	\end{figure*}

	After Encoding, we need to apply the encoded matrices to calculate psum, as shown in Tab.\ref{tab2}. First, we read the column index of matrix A and row index of matrix B. If $col\_idx\_sA < row\_idx\_sB$, we move the $col\_idx\_sA$ to next column in the 2\textasciitilde3-th line. If $col\_idx\_sA < row\_idx\_sB$, we move the $row\_idx\_sB$ to next row in the 4\textasciitilde5-th line. If $col\_idx\_sA = row\_idx\_sB$, we can calculate the psum. First, for those groups with $group\_bitmap = 1$, we fetch the elements of $value\_sA$, $value\_sB$, and $col\_idx\_sB$ to calculate the psum value, $value\_psum$, and column index, $col\_idx\_psum$, in the 13\textasciitilde14-th line. Besides, we store the memory address of $value\_psum$ and $col\_idx\_psum$, $id\_psum$, to the $row\_idx\_sA$ segmant of $vc\_addr\_psum$ in the 15-th line. Additionally, we record the length of psum, $row\_len\_psum$, in the $row\_idx\_sA$ row in the 16-th line so that we can access the psum row-wise regularly. When all elements in current $col\_idx\_sA$ and $row\_idx\_sB$ are finished, we traverse the remainings and repeat the above operations. \\
	\indent\setlength{\parindent}{1em}An example of psum calculation based on IOHP is shown in Fig.\ref{fig5}, where the input matrix comes from Fig.\ref{fig3}. At $T_1$, due to $col\_idx\_sA = row\_idx\_sB = 0$, we read the $group\_bitmap\_sA$ and $group\_bitmap\_sB$, indicating that $group0$ of submatrix A, $group0$ and $group1$ of submatrix B are non-empty. Since the $col\_len\_sA$ and $row\_len\_sB$ are both $1$, we fetch the $<value\_sA\,row\_idx\_sA> = <a1, 0>$, $<value\_sB, col\_idx\_sB> = <b1, 1>$, and $<value\_sB, col\_idx\_sB> = <b2, 1>$ to calculate psum. Then, we move to the next column, $col\_idx\_sA = 1$ (also next row, $row\_idx\_sB = 1$) to execute similar operations at $T_2$. The whole process takes $5$ cycles. The bottom right corner shows the process of psum calculation from submatrix A of $group0$ and submatrix B of $group1$. The $value\_psum$ and $col\_idx\_psum$ are stored in the order of $T_1$\textasciitilde$T_4$. Assuming the maximum length of non-zero psum is $4$, the memory addresses of $value\_psum$ and $col\_idx\_psum$ in $row0$ are stored in $address0-3$ of $vc\_addr\_psum$ and $row1$ in $address4-7$. Besides, the $row\_len\_psum$ are stored in the order of row index. Based on the $vc\_addr\_psum$ and $row\_len\_psum$, we can access the psum of each row regularly.
	  
	\begin{table}[H]\centering
		\caption{Computing Flow of Psum}
		\label{tab2}
		\begin{tabular}{lll}
			\toprule
			\textbf{Input}: submatrix A of RP\_CSC: $value\_sA$, $row\_idx\_sA$, \\
							$col\_len\_sA$, $col\_idx\_sA$, $group\_bitmap\_sA$;\\
			submatrix B of CP\_CSR : $value\_sB$, $col\_idx\_sB$, \\
							$row\_len\_sB$, $row\_idx\_sB$, $group\_bitmap\_sB$.  \\
			\textbf{Output}: psum: $value\_psum$, $col_idx_psum$, \\
							$row\_len_psum$, $addr\_col\_psum$.\\
			\midrule
			\textbf{1}\hspace{0.3cm} $while(col\_id<col\_all\_len\_sA\ \&$ \\
			\textbf{ }\hspace{1.3cm}		$row\_id<row\_all\_len\_sB)$\\
			\textbf{2}\hspace{0.6cm}	$if(col\_idx\_sA<row\_idx\_sB)$\\
			\textbf{3}\hspace{0.9cm}		$ena\_gb = 1,\ enb\_gb = 0$\\
			\textbf{4}\hspace{0.6cm}	$else if(col\_idx\_sA[col\_id]<row\_idx\_SB[row\_id])$\\
			\textbf{5}\hspace{0.9cm}		$ena\_gb = 0,\ enb\_gb\ = 1$\\
			\textbf{6}\hspace{0.6cm}	$else$\\
			\textbf{7}\hspace{0.9cm}		$ena\_gb = 1,\ enb\_gb = 1$\\
			\textbf{8}\hspace{0.6cm}	$for(g=0;g<G_{NA};g=g+1)$\\
			\textbf{9}\hspace{0.9cm}		$for(h=0;h<G_{NB};h=h+1)$\\
			\textbf{10}\hspace{1.1cm}			$if(ena\_gb == 1\ \&\ group\_bitmap\_sA[g]==1$ \\
			\textbf{}\hspace{1.3cm}				$\&\ enb\_gb ==1\ and\ group\_bitmap\_sB[h]==1)$\\
			\textbf{11}\hspace{1.4cm}				$for(i=0;i<col\_len\_sA[g];i=i+1)$\\
			\textbf{12}\hspace{1.7cm}					$for(j=0;j<row\_len\_sB[h];j=j+1)$\\			
			\textbf{13}\hspace{2.0cm}						$value\_psum[g][h] = value\_sA[g]*$\\
			\textbf{  }\hspace{4.9cm}											 $value\_sB[h]$\\
			\textbf{14}\hspace{2.0cm}						$col\_idx\_psum[g][h] = col\_idx\_sB[h]$\\
			\textbf{15}\hspace{2.0cm}						$vc\_addr\_psum[g][h][row\_idx\_sA[g]] = $\\
			\textbf{  }\hspace{5.2cm}											 $  id\_psum[g][h]$\\
			\textbf{16}\hspace{2.0cm}						$row\_len\_psum[g][h][row\_idx\_sA[g]]++ $\\
			\textbf{17}\hspace{2.0cm}						$id\_psum[g][h]++$\\
			\textbf{18}\hspace{0.6cm}	$col\_id += ena\_gb,\ row\_id += enb\_gb$\\
			\bottomrule
		\end{tabular}
	\end{table}

	\subsubsection{Address Mapping}
	After calculating all Psums, we need to sort and accumulate the irregular psums to obtain the output matrix C as shown in Tab.\ref{tab3}. For the $M\_t \times N\_t$ psum matrices, we traverse them row-wise in the 3-th line. First, we read the $row\_len\_psum$ in the 4-th line. If $row\_len\_psum>0$, we fetch the memory address of $col\_idx\_psum$, $vc\_addr\_psum$,  in the 5\textasciitilde6-th line. Then, we read the $col\_idx\_psum$ based on $vc\_addr\_psum$ and sort them in the current row in the 7\textasciitilde8-th line. Based on the sorted memory address, $sorted\_vc\_addr\_psum$, we read the $value\_psum$ and accumulate the adjacent psum with the same $col\_idx\_psum$ in the 9-th line. After that, we obtain the final result of matrix C, $<value\_C, row\_idx\_C, col\_idx\_C>$ in the 10-th line. When all psum are traversed, we can encode them with the standard CSC or CSR format in the 11-th line. 
	\begin{table}[H]\centering
		\caption{Computing Flow of Address Mapping}
		\label{tab3}
		\begin{tabular}{lll}
			\toprule
			\textbf{Input}: psum: $value\_psum$, $col\_idx\_psum$, \\
							$row\_len\_psum$, $vc\_addr\_psum$.\\
			\textbf{Output}: Matrix C of CSC (or CSC) format. \\
			\midrule
			\textbf{1}\hspace{0.3cm} $for(g=0;g<G_{NA};g=g+1)$\\
			\textbf{2}\hspace{0.6cm} 	$for(h=0;h<G_{NB};h=h+1)$\\
			\textbf{3}\hspace{0.9cm} 		$for(i=0;i<M_t;i=i+1)$\\
			\textbf{4}\hspace{1.2cm} 			$if(row\_len\_tmp[g][h]>0)$\\
			\textbf{5}\hspace{1.5cm} 				$vc\_addr\_tmp = vc\_addr\_psum[g][h]$ \\
			\textbf{6}\hspace{1.5cm} 				$col\_idx\_tmp = col\_idx\_psum[g][h]$\\
			\textbf{7}\hspace{1.5cm} 				$(sorted\_vc\_addr\_tmp,sorted\_col\_idx\_tmp)$\\
			\textbf{ }\hspace{1.8cm}				  $ = sort(vc\_addr\_tmp,col\_idx\_tmp)$\\
			\textbf{8}\hspace{1.5cm} 				$value\_C\_tmp = value\_psum[g][h]$ \\
			\textbf{9}\hspace{1.5cm} 				$accum(value\_C\_tmp, sorted\_col\_idx\_tmp)$\\
			\textbf{10}\hspace{1.4cm} 				$(value\_C, row\_idx\_C,col\_idx\_C)$\\
			\textbf{ }\hspace{1.6cm}				$ = (value\_C\_tmp,i,sorted\_col\_idx\_tmp)$\\
			\textbf{11}\hspace{0.3cm} $Encode\_CSC(value\_C, row\_idx\_C, col\_idx\_C)$\\
			\bottomrule
		\end{tabular}
	\end{table}
	\begin{figure}[H]
		\centering
		\includegraphics[width=0.99\linewidth]{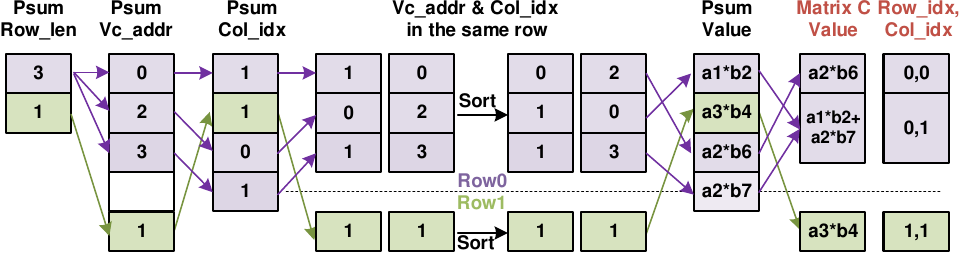}
		\caption{An example of address mapping.}
		\label{fig6}
	\end{figure} 
	\indent\setlength{\parindent}{1em}An example of address mapping is shown in Fig.\ref{fig6}. The $row\_len\_psum$ of $row0$ and $row1$ are $3$ and $1$ respectively. For $row0$, we fetch the $col\_idx\_psum = <1, 0, 1>$ based on the $vc\_addr\_psum = <0, 2, 3>$. Then, we sort the $col\_idx\_psum$ with corresponding $vc\_addr\_psum$. After that, we fetch the $value\_psum = <a1*b2, a3*b4, a2*b6>$ based on the sorted $vc\_addr\_psum = <2, 0, 3>$. The $value\_psum = a1*b2$ and $a2*b6$ are accumulated due to the same $col\_idx\_psum$.  Finally, we obtain the value and location of matrix C, $<value\_C, row\_idx\_C, col\_idx\_C> = <(a2*b6, 0, 0), (a1*b2+a2*b7, 0, 1), (a2*b6, 1, 1)>$. For $row1$, we excute the similar operations, the $<value\_C, row\_idx\_C, col\_idx\_C> = <(a3*b4, 1, 1)>$.  
	 
	\subsection{Computing flow for SDMM}

\begin{figure}[H]
		\centering
		\includegraphics[width=0.99\linewidth]{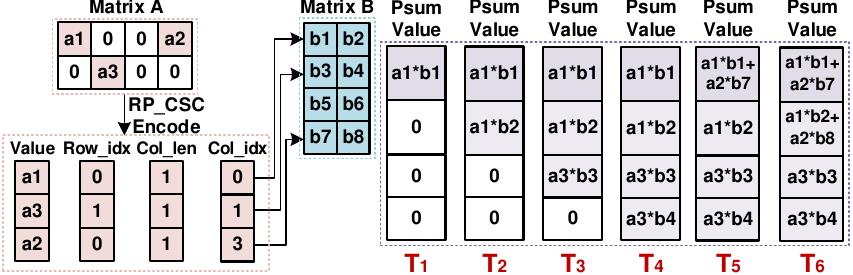}
		\caption{An example of SDMM based on IOHP.}
		\label{fig7}
	\end{figure}

	For SDMM, we can simplify the flow of IOHP, as shown in Tab.\ref{tab4}. First, we only need to encode matrix A with RP-CSC format. Then, we fetch the value of matrix B, $value\_B$, according to the column index of matrix A, $col\_idx\_sA$, in the 2-th line. Thus, we calculate column-wise. Since the psum matrix is dense, the location of each element is also regular. For those groups with $group\_bitmap = 1$,  we calculate the $value\_psum$ and accumulate with the old psum in the 9-th line. When all elements are finished, we obtain a dense matrix C.    
	\begin{table}[H]\centering
		\caption{Computing flow of SDMM based on IOHP}
		\label{tab4}
		\begin{tabular}{lll}
			\toprule
			\textbf{Input}: Matrix A of CSC format: $value\_A$, $row\_idx\_A$, \\
							 $col\_ptr\_A$; Matrix B.\\
			\textbf{Output}: Matrix C. \\
			\midrule
			\textbf{1}\hspace{0.3cm} $Encode\_RP\_CSC(value\_A,row\_idx\_A,col\_ptr\_A)$\\
			\textbf{2}\hspace{0.3cm} $value\_sB = value\_B[col\_idx\_sA]$ \\
			\textbf{3}\hspace{0.3cm} $while(col\_id<col\_all\_len\_sA)$\\
			\textbf{4}\hspace{0.6cm} 	$for(g=0;g<G_{NA};g=g+1)$\\
			\textbf{5}\hspace{0.9cm} 		$for(h=0;h<G_{NB};h=h+1)$\\
			\textbf{6}\hspace{1.2cm} 			$if(group\_bitmap\_sA[g]==1)$\\
			\textbf{7}\hspace{1.5cm} 				$for(i=0;i<col\_len\_sA[g];i=i+1)$\\
			\textbf{8}\hspace{1.8cm} 					$for(j=0;j<N_t;j=j+1)$\\
			\textbf{9}\hspace{2.1cm} 						$value\_psum[g][h] +=$ \\
			\textbf{ }\hspace{2.1cm} 						$value\_sA[g]*value\_sB[h]$\\
			\textbf{10}\hspace{0.5cm} 	$col\_id++$\\
			\bottomrule
		\end{tabular}
	\end{table}
	\indent\setlength{\parindent}{1em}Fig.\ref{fig7} shows an example of SDMM based on IOPS. First, the matrix A is encoded with RP-CSC format. Then, we fetch the $row0, row1, row3$ of matrix B based on the $col\_idx\_sA = <0,1,3>$. After that, we calculate the psum to obtain matrix C. At $T_1$, we calculate the $value\_psum = a1*b1$ and store it to $address = 0$. At $T_5$, we get $value\_psum = a2*b7$ and accumulate with $value\_psum = a1*b1$ of $address = 0$. The whole process take $6$ cycles and obtain a $2\times2$ dense matrix C. 
	\section{Architecture}
	
	\subsection{Top-level Architecture}
\begin{figure}[H]
		\centering
		\includegraphics[width=0.65\linewidth]{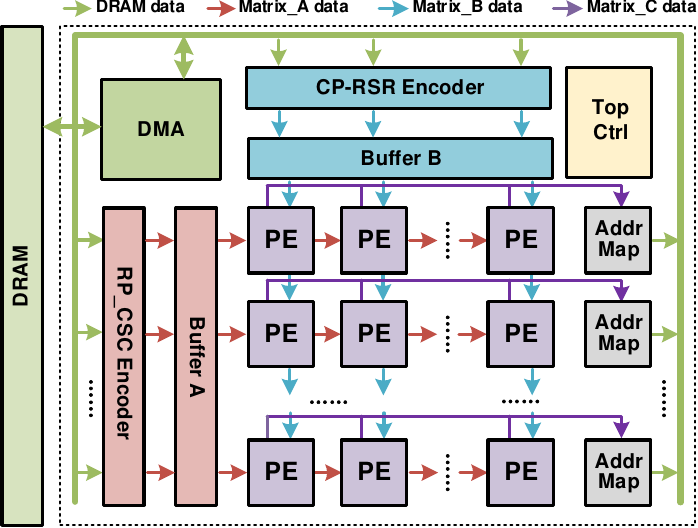}
		\caption{Top-level architecture of IOPS.}
		\label{fig8}
	\end{figure}

	To implement the IOHP approach, we design a hardware architecture, IOPS, as shown in Fig.\ref{fig8}. It mainly consists of RP-CSC encoder, CP-CSR encoder, input buffers for matrices A and B (Buffer A, B), a PE array, address mapping modules (Addr Map), a top controller and a direct memory access (DMA) module. Initially, we load the orignal input matrices from DRAM to the encoders through DMA. Then, we encode the input matrices with RP-CSC and CP-CSR format and transfer the encoded matrices to buffer A and B. After that, the encoded matrices are fetched to PE array to calculate psum based on IOHP. Finally, the psum are sorted and accumulated regularly to obtain the output matrix C in the address mapping module, which also can be encoded with CSC or CSR format. To improve performance, we organize three major stages, encoding input matrices with RP-CSC and CP-CSR format, calculating psum based on IOHP, and accumulating psum through address mapping, as a pipeline structure. Besides, we can switch the computing mode between SSMM and SDMM to adpat different applications. 

	\subsection{PE array and Address Mapping Module}
\begin{figure}[H]
		\centering
		\includegraphics[width=0.99\linewidth]{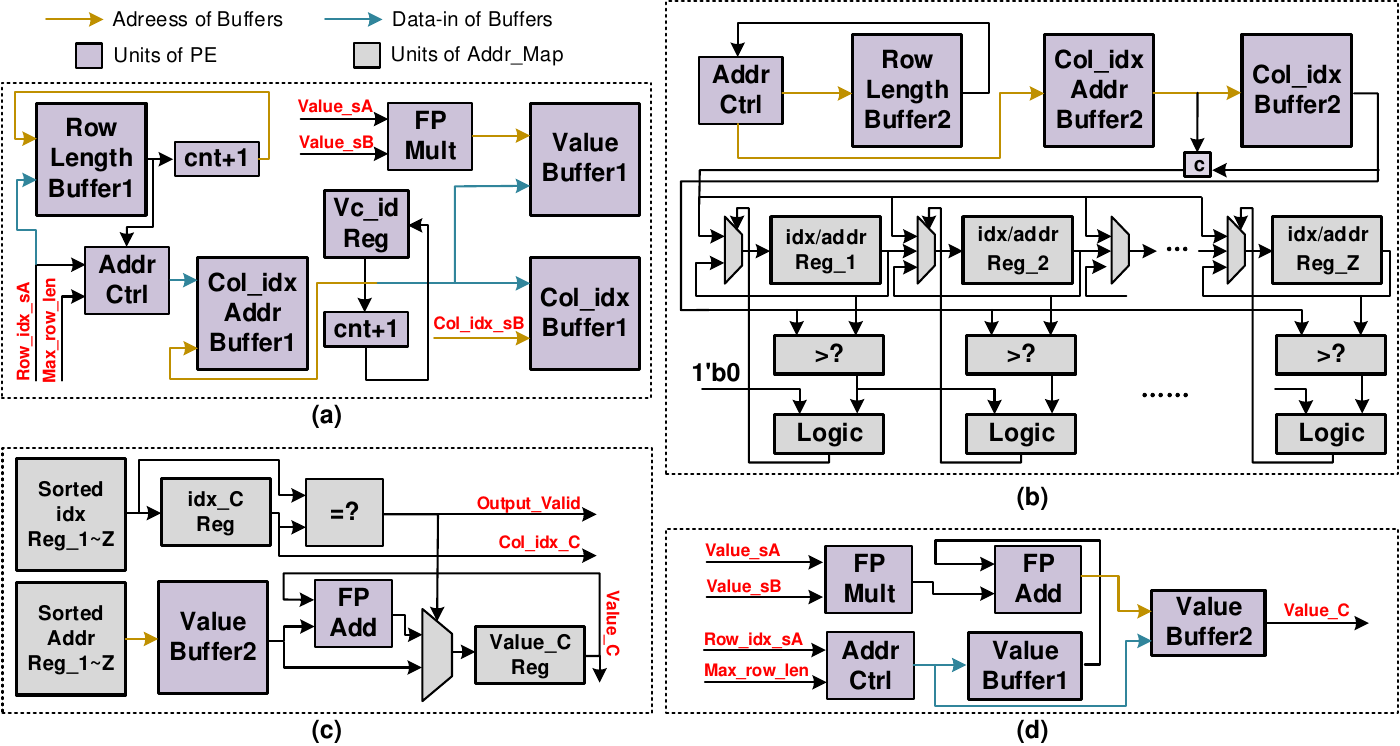}
		\caption{PE and address mapping unit. (a) Calculating psum. (b) Sorting the location of psum. (c) Accumulating psum. (d) Computing mode of SDMM.}
		\label{fig9}
	\end{figure}

	The PE array, the core of IOPS, is in charge of SpMM, which consists of $G_{NA} \times G_{NB}$ PE units. Each PE consists of a float-point multiplier and adder (FP-mult and FP-Add), two sets of row\_len buffers, value buffers, vc\_addr buffers. and col\_idx buffers. One set of buffers are appiled to store the data during calculating psum as shown in Fig.\ref{fig9}(a). The other set of buffers are used for fetching psum to obtain the output matrix collaborating with address mapping module as shown in Fig.\ref{fig9}(b) and Fig.\ref{fig9}(c). \\
	\indent\setlength{\parindent}{1em}To sort the location of psum, we equip two sequences of shift registers (idx/addr reg), comparators ($>?$), and multiplexer logic as shown in Fig.\ref{fig9}(b). First, we fetch a $col\_idx\_psum$ and a $vc\_addr\_psum$ from buffers and compare the $col\_idx\_psum$ with all registers. If the $col\_idx\_psum$ is larger than the current register, we keep the original value. If the $col\_idx\_psum$ is smaller than the current register while larger than the former one, the $col\_idx\_psum$ and $vc\_addr\_psum$ are inserted to the location of current register and the remaining registers shift right. After sorting, we read the $value\_psum$ one by one from value buffer based on the sorted $vc\_addr\_psum$ (sorted addr reg) as shown in Fig.\ref{fig9}(c). Meanwhile, the sorted $col\_idx\_psum$ (sorted idx reg) are fetched serially to detect the equality of the current and the former data. If equal, the $value\_psum$ is accumulated with the former one. Otherwise, the former $value\_psum$ is output as a final result of matrix C. \\
	\indent\setlength{\parindent}{1em}For the computing mode of SDMM, since the buffers corresponding to address mapping operation are idle, they can be merged with value buffers for a larger storage as shown in Fig.\ref{fig9}(d). With more on-chip buffers, the size of submatrix becomes larger, obviously reducing the tiling blocks and lowering DRAM access further. Finally, we apply two value buffers to accumulate psum with a ping-pong buffer structure.
	   
	\section{Adaptive Partition Strategy}
	
\begin{figure}[H]
		\centering
		\includegraphics[width=0.99\linewidth]{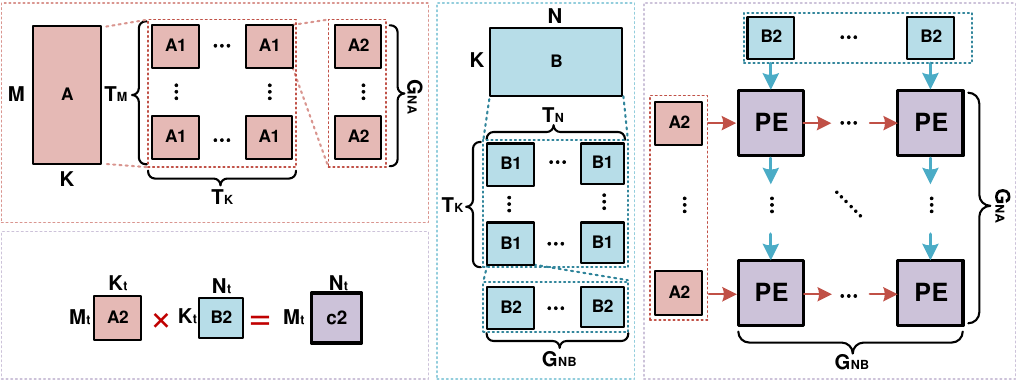}
		\caption{Adaptive partition strategy.}
		\label{fig10}
	\end{figure}

	\indent\setlength{\parindent}{1em}To reduce the DRAM access of overall architecture, we propose an adaptive partition strategy. Based on IOHP, we partition the matrix A and matrix B as shown in Fig.\ref{fig10}. The specific partitioning parameters are shown in Tab.\ref{tab5}. Matrix A and B are partitioned into $T_{M}\times T_K$ blocks of $A_1$ and $T_K \times T_N$ blocks of $B_1$, respectively. Each $A_1$ contains $G_{NA}$ groups of $A_2$ and each $B_1$ contains $G_{NB}$ groups of $B_2$, where each $A_2$ and $B_2$ are shared by the PEs in the same row and column, respectively. Each PE performs an outer product between $A_2$ and $B_2$, obtaining an output submatrix $C_2$ with the size of $M_t\times N_t$. 
	\begin{table}[H]\centering
		\caption{Parameters of Data Partition}
		\label{tab5}
		\begin{tabular}{|c|c|}
			\hline
			Parameter & Explanation                 \\ \hline
			$C_{psum}$       & capacity of psum buffer               \\ \hline
			$C_{A}$/$C_{B}$       & capacity of input buffer A/B              \\ \hline
			$M_{A}$/$M_{B}$       & storage size of matrix A/B                   \\ \hline
			$M\times K$        &  size of matrix A                     \\ \hline
			$K\times N$       &  size of matrix B             \\ \hline
			$D_{A}$/$D_{B}$       & density of matrix A/B             \\ \hline
			$G_{NA}/G_{NB}$       &  group \# of matrix A2/B2                    \\ \hline
			$ R_x $			&  reserved storage ratio of buffer \\ \hline
			$T_{M}$        & tiling \# of matrix A row  \\ \hline
			$T_{K}$        & tiling \# of matrix A column (matrix B row)  \\ \hline
			$T_{N}$       & tiling \# of matrix B column  \\ \hline
			$M_t\times K_t$   &  size of submatrix A2      \\ \hline
			$K_t\times N_t$   &  size of submatrix B2     \\ \hline
		\end{tabular}
	\end{table}
	\indent\setlength{\parindent}{1em}According to the partition method in Fig.\ref{fig10} and Tab.\ref{tab5}, we need to solve a set of tiling parameters to minimize the DRAM access with limited buffer capacity. The specific algorithm is shown in Tab.\ref{tab6}. Assuming an uniform distribution of sparsity in the input matrix, the $M_t$, $K_t$ and $N_t$ should satisfy constraints $(1)-(3)$, which are the capacity limitations of buffer A, B, and psum, respectively. Generally, we reserve $R_x$ free storage space due to the deviation between theory and reality. If it still exceeds the capacity of on-chip buffers, we will stop computing and load the psum of overflowed buffer into DRAM or read the rest input matrices from DRAM, which rarely happens. If the input buffer is capacious enough to store the submatrix A of $M_t\times K$ or B of $K\times N_t$, we can reuse matrix A first (RAF) or B first (RBF) to avoid repeat access, as shown in $DrA_0$ and $DrA_1$. Otherwise, we reuse matrix A and B equally (RABE) and the DRAM access is calculated as $DrA_2$. Finally, since the solution space of tiling parameter is limited, we apply the exhaustion method to solve it for minimizing DRAM access.
	
		\begin{table}[H]\centering
		\caption{Strategy of Data Partition}
		\label{tab6}
		\begin{tabular}{c|c}
			\hline
			Input & $C_{psum}$, $C_{A}$, $C_{B}$, $M_{A}$, $M_{B}$, $R_x$\\
				  &	$M$, $K$, $N$, $D_A$, $D_B$, $G_{NA}$, $G_{NB}$ \\ \hline
			Output & $T_{M}$, $T_{K}$, $T_{N}$, $M_t$, $K_t$, $N_t$ \\ \hline
			 			& $(1)\ M_{t}\times K_{t} \times D_A< C_{A} \times R_x$ \\ 
			Constraint	& $(2)\ K_{t}\times N_{t}  \times D_B < C_{B} \times R_x$ \\ 
						& $(3)\ M_{t} \times D_A \times  N_{t} \times D_B \times K_t < C_{psum} \times R_x$ \\ \hline
			Extra		& $(a)\ M_{t}\times K \times D_A< C_{A} \times R_x$ \\
			Condition	& $(b)\ N_{t}\times K \times D_B< C_{B} \times R_x $ \\	\hline	
					 & $Min(DrA_0, DrA_1, DrA_2), where$ \\
			Target	&$RAF: DrA_0 = M_{B}\times T_{M}+M_{A}, if (a) $ \\
 					& $RBF: DrA_1 = M_{A}\times T_{N}+M_{B}$,  $if (b)$ \\
					&$RABE: DrA_2 = M_{A}\times T_{N}+M_{B}\times T_{M}$ \\ \hline
			Algorithm & exhaustion method \\ \hline
		\end{tabular}
	\end{table}

	\section{Experiment}
	
	\subsection{Implementation}
	IOPS implementation: To measure the area, power, and performance of hardware, we implemented IOPS in Verilog and conducted a functional simulation on Cadence Xcelium 2019.2. Then, the Verilog code was synthesized with Synopsys Design Compiler under the 65-nm technology. The on-chip SRAM buffers are generated by Memory Compiler and the power is calculated by Synopsys Prime Suite 2020.09.  \\
	\indent\setlength{\parindent}{1em} The PE array is $8\times8$, achieving a peak throughput of $64\times0.8GHz = 51.2 GOP/s$. For original input matrix, the value, location, and data length are encoded in 64-bit (double float), 32-bit, and 16-bit, where these data of RP-CSC and CP-CSR format are encoded in 64-bit, 16-bit, and 8-bit, respectively. To balance performance and hardware overhead, the sizes of each data buffer are set as shown in Tab.\ref{tab7}, where all of them are doubled as a ping-pong buffer structure. In SDMM mode, the buffer of col\_idx, row\_len, and vr\_addr can be merged to an additional $256\times64b$ value buffer, which expands the on-chip storage and reduces the DRAM access. For the area and power breakdown of each module,  PE array consumes 58.3\% area and 65.4\% power for calculating SpMM and storing psum data. Input buffers mainly store input matrices of RP-CSC and CP-CSR format, accounting for  32.2\% area and 22.8\% power. The Encoders provide a buffer for the original input matrices to encode them, with 7.2\% area and 7.2\% power consumption. Other parts such as controller and address mapping module are in charge of dataflow control and psum regularization, occupying only 2.4\% area and 4.6\% power. 
	\begin{table}[H]\centering
		\caption{Area and Power of Each Module}
		\label{tab7}
		\begin{tabular}{|c|c|c|c|}
			\hline
			Module & Area(mm2)                                              & Power(W)                                                & \multicolumn{1}{c|}{Buffer Size}                                                                                                            \\ \hline
			PE Array  & \begin{tabular}[c]{@{}c@{}}9.8\\ (58.2\%)\end{tabular} & \begin{tabular}[c]{@{}c@{}}2.09\\ (65.4\%)\end{tabular} & \multicolumn{1}{c|}{\begin{tabular}[c]{@{}c@{}}value: 256x64b,\\  col\_idx: 256x16b, \\ row\_len: 2x256x8b, \\ vr\_addr:4x256x8b\end{tabular}} \\ \hline
			Buffer A  & \begin{tabular}[c]{@{}c@{}}2.7\\ (16.1\%)\end{tabular} & \begin{tabular}[c]{@{}c@{}}0.36\\ (11.4\%)\end{tabular} & \begin{tabular}[c]{@{}l@{}}value: 2Kx64b, \\ row\_idx: 2Kx16b, \\ col\_len: 2Kx8b, \\ col\_idx: 8Kx32b,\\ bitmap:8Kx16b\end{tabular}        \\ \hline
			Buffer B  & \begin{tabular}[c]{@{}c@{}}2.7\\ (16.1\%)\end{tabular} & \begin{tabular}[c]{@{}c@{}}0.36\\ (11.4\%)\end{tabular} & \begin{tabular}[c]{@{}l@{}}value: 2Kx64b, \\ col\_idx: 2Kx16b, \\ row\_len: 2Kx8b, \\ row\_idx: 8Kx32b,\\  bitmap:8Kx16b\end{tabular}       \\ \hline
			Encoder A & \begin{tabular}[c]{@{}c@{}}0.6\\ (3.6\%)\end{tabular}  & \begin{tabular}[c]{@{}c@{}}0.12\\ (3.6\%)\end{tabular}  & \begin{tabular}[c]{@{}l@{}}value: 256x64b, \\ row\_idx: 256x32b, \\ col\_ptr: 256x16b\end{tabular}                                          \\ \hline
			Encoder B & \begin{tabular}[c]{@{}c@{}}0.6\\ (3.6\%)\end{tabular}  & \begin{tabular}[c]{@{}c@{}}0.12\\ (3.6\%)\end{tabular}  & \begin{tabular}[c]{@{}l@{}}value: 256x64b, \\ col\_idx: 256x32b, \\ row\_ptr: 256x16b\end{tabular}                                          \\ \hline
			Others    & \begin{tabular}[c]{@{}c@{}}0.4\\ (2.4\%)\end{tabular}  & \begin{tabular}[c]{@{}c@{}}0.15\\ (4.6\%)\end{tabular}  & -                                                                                                                                           \\ \hline
			Total     & \begin{tabular}[c]{@{}c@{}}16.8\\ (100\%)\end{tabular} & \begin{tabular}[c]{@{}c@{}}3.2\\ (100\%)\end{tabular}   & -                                                                                                                                           \\ \hline
		\end{tabular}
	\end{table}
	\indent\setlength{\parindent}{1em}To validate the superiority of this work, we set some baseline accelerators as shown in Tab.\ref{tab8}, including inner and outer product based architectures, as well as SSMM and SDMM accelerators. The benchmark is shown in Tab.\ref{tab9}. For SSMM, we testified with some real world sparse matrices in SNAP datasets \cite{snapnets}, where the SSMM is calculated by the adjacent matrix ($A$) and feature matrix ($X$). For SDMM, we testified with a 2-layer GCN model \cite{kipf2017semisupervised}  on the Cora\cite{Cora}, Citeseer\cite{10.1145/276675.276685}, Pubmed\cite{Sen2008CollectiveCI}, Nell\cite{10.5555/2898607.2898816}, and Reddit\cite{10.1145/2783258.2783417} datasets, where the SDMM is conducted by $A((A(XW_1))W_2)$ and the weight matrices, $W_1, W_2$, are dense.

	\begin{table*}\centering
		\caption{Overall Comparison of Each Accelerator}
		\label{tab8}
		\begin{tabular}{|c|c|c|c|c|c|c|c|c|}
			\hline
			Accelerators & \begin{tabular}[c]{@{}c@{}}Technology\\  (nm)\end{tabular} & \begin{tabular}[c]{@{}c@{}}Technology Scalar \\ (Area/Power)\end{tabular} & \begin{tabular}[c]{@{}c@{}}PE Array\\ Size\end{tabular} & \begin{tabular}[c]{@{}c@{}}Area \\ (mm2)\end{tabular} & \begin{tabular}[c]{@{}c@{}}Frequency \\ (MHz)\end{tabular} & \begin{tabular}[c]{@{}c@{}}Power \\ (W)\end{tabular} & \begin{tabular}[c]{@{}c@{}}Computing\\ Approach\end{tabular}   & \begin{tabular}[c]{@{}c@{}}Expertise\\ Domains\end{tabular} \\ \hline
			SPOTS \cite{ 10.1145/3532863}       & 45                                                         & 2x/1.6x                                                                   & 4x128                                                   & 8.6                                                   & 500                                                        & 1.1                                                  & Inner Product                                                  & DDMM                                                        \\ \hline
			HyGCN \cite{9065592}       & 12                                                         & 29.3/23.4                                                                 & \begin{tabular}[c]{@{}c@{}}8x4x128\\ 32x16\end{tabular} & 7.8                                                   & 1000                                                       & 6.7                                                  & Inner Product                                                  & \begin{tabular}[c]{@{}c@{}}DDMM\\ SSMM\end{tabular}         \\ \hline
			GCNAX \cite{9407104}       & 40                                                         & 2.6x/2.1x                                                                 & 1x16                                                    & 6.5                                                   & 1000                                                       & 0.37                                                 & Outer Product                                                  & SDMM                                                        \\ \hline
			OuterSPACE \cite{8327050}  & 32                                                         & 4.1x/3.3x                                                                 & 16x16                                                   & 87                                                    & 1000                                                       & 12.39                                                & Outer Product                                                  & SSMM                                                        \\ \hline
			SpArch \cite{9065428}      & 40                                                         & 2.6x/2.1x                                                                 & 16x16                                                   & 28.49                                                 & 1000                                                       & 9.26                                                 & Outer Product                                                  & SSMM                                                        \\ \hline
			IOPS         & 65                                                         & 1x/1x                                                                     & 8x8                                                     & 16.8                                                  & 800                                                        & 3.2                                                  & \begin{tabular}[c]{@{}c@{}}Inner-Outer\\  Product\end{tabular} & \begin{tabular}[c]{@{}c@{}}SSMM\\ SDMM\end{tabular}         \\ \hline
		\end{tabular}
	\end{table*}
	
	\begin{table}[H]\centering
		\caption{Matrix Density and Demension of Each Dataset}
		\label{tab9}
		\begin{tabular}{|c|c|c|c|}
			\hline
			Dataset  & \begin{tabular}[c]{@{}c@{}}Vertex (A),\\ Feature (X)\end{tabular} & \begin{tabular}[c]{@{}c@{}}Density of\\ A, X\end{tabular}   & \begin{tabular}[c]{@{}c@{}}Weight matrix\\ W1,W2\end{tabular}          \\ \hline
			cora     & \begin{tabular}[c]{@{}c@{}}2708,\\ 1433\end{tabular}              & \begin{tabular}[c]{@{}c@{}}0.14\%,\\ 1.27\%\end{tabular}    & \begin{tabular}[c]{@{}c@{}}{[}1433,16{]}, \\ {[}16,7{]}\end{tabular}   \\ \hline
			citeseer & \begin{tabular}[c]{@{}c@{}}3327,\\ 3703\end{tabular}              & \begin{tabular}[c]{@{}c@{}}0.083\%,\\ 0.85\%\end{tabular}   & \begin{tabular}[c]{@{}c@{}}{[}3703,16{]}, \\ {[}16,6{]}\end{tabular}   \\ \hline
			pumbed   & \begin{tabular}[c]{@{}c@{}}19717,\\ 500\end{tabular}              & \begin{tabular}[c]{@{}c@{}}0.023\%,\\ 10\%\end{tabular}     & \begin{tabular}[c]{@{}c@{}}{[}500,16{]},\\ {[}16,3{]}\end{tabular}     \\ \hline
			nell     & \begin{tabular}[c]{@{}c@{}}169343,\\ 61278\end{tabular}           & \begin{tabular}[c]{@{}c@{}}0.0041\%,\\ 0.011\%\end{tabular} & \begin{tabular}[c]{@{}c@{}}{[}61278,64{]},\\ {[}64,186{]}\end{tabular} \\ \hline
			Reddit   & \begin{tabular}[c]{@{}c@{}}232965,\\ 602\end{tabular}             & \begin{tabular}[c]{@{}c@{}}0.021\%,\\ 51.6\%\end{tabular}   & \begin{tabular}[c]{@{}c@{}}{[}602,64{]},\\ {[}64,41{]}\end{tabular}    \\ \hline
			Facebook & \begin{tabular}[c]{@{}c@{}}22470,\\ 4714\end{tabular}             & \begin{tabular}[c]{@{}c@{}}0.034\%\\ 0.3\%\end{tabular}     & -                                                                      \\ \hline
			Deezer   & \begin{tabular}[c]{@{}c@{}}28281,\\ 30978\end{tabular}            & \begin{tabular}[c]{@{}c@{}}0.011\%\\ 0.11\%\end{tabular}    & -                                                                      \\ \hline
			Twitch   & \begin{tabular}[c]{@{}c@{}}9498,\\ 2514\end{tabular}              & \begin{tabular}[c]{@{}c@{}}0.17\%,\\ 0.81\%\end{tabular}    & -                                                                      \\ \hline
			Wiki     & \begin{tabular}[c]{@{}c@{}}11631,\\ 13183\end{tabular}            & \begin{tabular}[c]{@{}c@{}}0.25\%,\\ 0.84\%\end{tabular}    & -                                                                      \\ \hline
			Github   & \begin{tabular}[c]{@{}c@{}}37700,\\ 4005\end{tabular}             & \begin{tabular}[c]{@{}c@{}}0.04\%,\\ 0.46\%\end{tabular}    & -                                                                      \\ \hline
		\end{tabular}
	\end{table}

	\subsection{Performance}
\begin{figure}[H]
		\centering
		\includegraphics[width=0.99\linewidth]{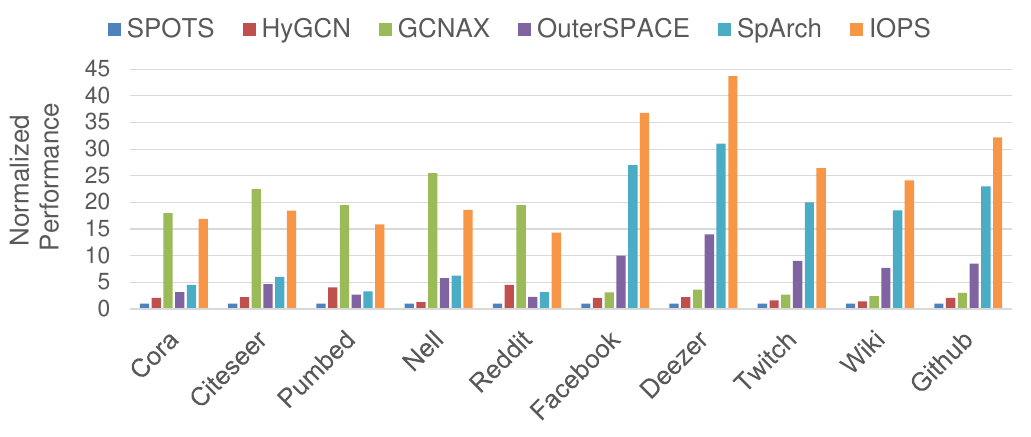}
		\caption{Performance comparison with SPOTS, HyGCN, GCNAX, OuterSPACE, and SpArch.}
		\label{fig11}
	\end{figure}

	To demonstrate the superiority of IOPS, we compared the performance with the SPOTS \cite{ 10.1145/3532863}, HyGCN \cite{9065592}, GCNAX \cite{9407104}, OuterSPACE \cite{8327050}, and SpArch \cite{9065428}, achieving $14.3\times$\textasciitilde$43.7\times$, $3.2\times$\textasciitilde$19.8\times$, $0.72\times$\textasciitilde$12.1\times$, $2.9\times$\textasciitilde$6.4\times$, and $1.3\times$\textasciitilde$4.8\times$, respectively, as shown in Fig.\ref{fig11}. To eliminate the influence of hardware scale and technology node, the normalized performance, $P_{N}$, is calculated as $R_{Pr} / R_{Pt}$, where $R_{Pr}$ and $R_{Pt}$ are the ratios of real performance and top performance of IOPS to other accelerators, respectively.  
	\indent\setlength{\parindent}{1em}For SPOTS, it reuses the input matrices across PEs array based on inner product and improves the performance by removing the rows and columns with all zeros. However, this approach can only skip minority zero computing due to the mismatched index between input matrix A and B. Compared with SPOTs, we conduct the outer product in each PE to avoid the problem. Thus, our performance is $14.3\times$\textasciitilde$43.7\times$ higher than SPOTS.\\
	\indent\setlength{\parindent}{1em}For HyGCN, it integrates the SIMD cores for SSMM and systolic arrays for DDMM. The window shrinking method of HyGCN dynamically eliminates the sparsity of the inner-product based computing kernel. For SSMM benchmark, the DDMM computing engine works like SPOTS in the best case. For SDMM benchmark, the execution order of $(A\times X)\times W$ causes increment of operation compared with $A\times(X\times W)$. Besides, the workloads between two computing engines can be imbalanced with the variation of input matrix size and density. Compared with HyGCN, we designed an unified architecture with SSMM and SDMM mode, making full use of the computing and memory resources. Thus, we achieve $3.2\times$\textasciitilde$19.8\times$ performance. \\
	\indent\setlength{\parindent}{1em}For GCNAX, it applies a row remapping strategy to accumulate psum more regularly, which achieves great performance in SDMM. However, it only exploits the sparsity of one input matrix, which gains less benefits in SSMM. Thus, the performance of GCNAX is $1.1\times$\textasciitilde$1.4\times$ higher than ours in the SDMM benchmark because the computing flow of SDMM is similar with us but the workload of GCNAX is more balanced with a smaller PE array. If our PE array is scaled to the same size with GCNAX, we can achieve the same performance theoretically. For the SSMM benchmark, we achieve $9.8\times$\textasciitilde$12.1\times$ speedup in  due to sparsity exploiting of two matrices.  \\
	\indent\setlength{\parindent}{1em}For OuterSPACE, although it removes most of the zero element computations in the input matrix through outer product approach, the irregular distribution of non-zero elements in the psum matrix leads to longer accumulation delays. Compared with OuterSPACE, we applied an address mapping method to regularize the psum accumulating. Thus, our performance is $2.9\times$\textasciitilde$6.4\times$ higher than OuterSPACE. \\
	\indent\setlength{\parindent}{1em}For SpArch, it condenses the input matrices to reduce the psum matrix and speeds up the accumulation process with the merge tree. However, no data reusing among PEs leads to longer input data loading delays, which can be 
	deteriorative in SDMM. Compared with SpArch, the benefit of optimizing psum accumulation is similar to us, but we reuse the input martices based on IOHP. Thus, we achieve $1.3\times$\textasciitilde$4.8\times$ performance compared with SpArch.

	\subsection{Energy and Resource Efficiency}
\begin{figure}[H]
		\centering
		\includegraphics[width=0.99\linewidth]{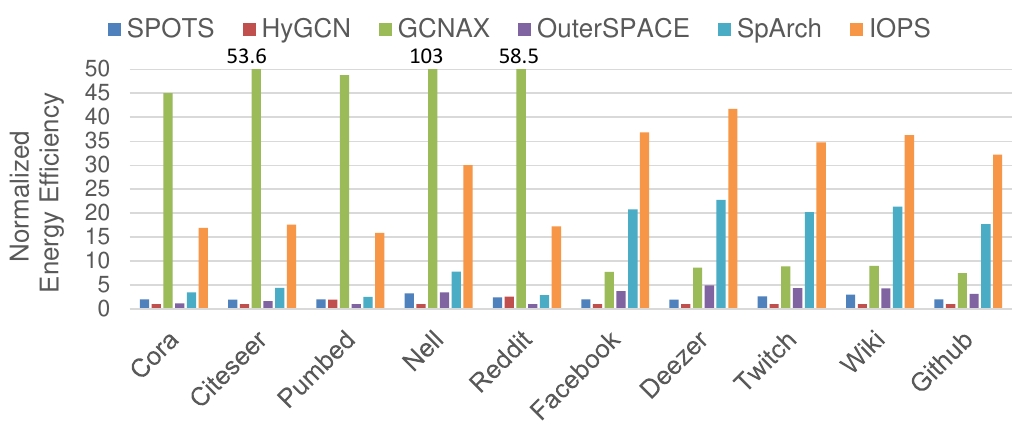}
		\caption{Energy efficiency comparison with SPOTS, HyGCN, GCNAX, OuterSPACE, and SpArch. }
		\label{fig12}
	\end{figure}

\begin{figure}[H]
		\centering
		\includegraphics[width=0.99\linewidth]{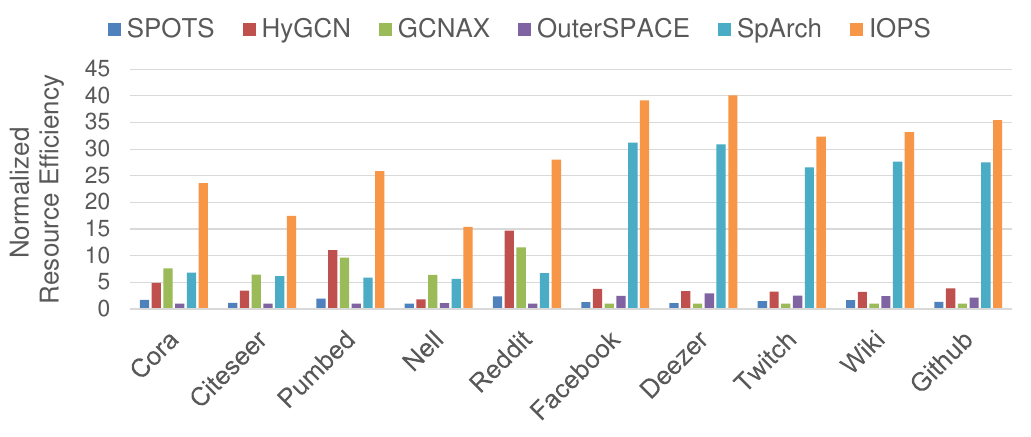}
		\caption{Resource efficiency comparison with SPOTS, HyGCN, GCNAX, OuterSPACE, and SpArch. }
		\label{fig13}
	\end{figure}

	To demonstrate the benefit of performance outweighs the area and power overhead in IOPS, we compared the energy and resource efficiency with SPOTS, HyGCN, GCNAX, OuterSPACE, and SpArch, achieving $7.2\times$\textasciitilde$21.9\times$, $6.7\times$\textasciitilde$41.7\times$,  $0.3\times$\textasciitilde$4.9\times$, $7.9\times$\textasciitilde$17.1\times$, and $1.7\times$\textasciitilde$6.3\times$ energy efficiency, respectively, as shown in Fig.\ref{fig12}, and achieving $11.9\times$\textasciitilde$36.3\times$, $1.9\times$\textasciitilde$11.9\times$, $2.4\times$\textasciitilde$40\times$, $12.9\times$\textasciitilde$28\times$, and $1.2\times$\textasciitilde$4.4\times$ resource efficiency, respectively, as shown in Fig.\ref{fig13}. The normalized energy efficiency, $E_N$, is calculated as $R_{Pr}$/($R_{w}\times s$), where $R_{w}$ is the ratio of power of IOPS to other accelerators, and $s$ is the power scalar \cite{7013055} among different technologies. To explain the difference of architecture design, $E_N$ can be calculated as $(R_{Pr}/R_{Pt})/(R_{W}/R_{Pt}\times s) = P_N\times E_{T}$, where the $E_{T}$ is the top energy efficiency, representing the power consumption per throughput. Similarly, the normalized resource efficiency, $R_{N}$, is calculated as $R_{N} = P_{N}\times R_{T}$, where the $R_{T}$ is the top resource efficiency, representing the area cost per throughput. \\
	\indent\setlength{\parindent}{1em}For SPTOS, the PE unit of systolic array only consists of one MAC unit and psum buffer, simplifying the structure and leaving out many operations compared with IOPS. However, the imbalanced row and column size of PE array induces more buffers for a higher bandwidth requirement. Besides, the data value of SPOTS are 16-bit fixed point, which needs to scale $4\times$ power and area to match the hardware of 64-bit float point. Thus, our top energy efficiency, $E_{T}$, and top resource efficiency, $R_T$, are $0.5\times$ and $0.83\times$ lower than SPOTS. Combining with the normalized performance, $P_N$, we achieve $7.2\times$\textasciitilde$21.9\times$ normalized energy efficiency, $E_N$, and $11.9\times$\textasciitilde$36.3\times$ normalized resource efficiency, $R_N$, respectively. 
	\indent\setlength{\parindent}{1em}For HyGCN, the DDMM engine based on systolic arrays can achieve higher throughput with lower area, but also consumes more power due to the denser computing units. Besides, the interface between the SDMM and DDMM engine occupies 35\% area and 20\% power for establishing the pipeline, which is an extra module compared with IOPS.  Additionaly, the 32-bit fixed point data of HyGCN scales $2\times$ power and area to match with double float point. Therefore, our $R_T$ is $0.7\times$ lower than HyGCN but our $E_{T}$ is $2.1\times$ higher than it. Finally, our $E_N$ and $R_N$ are $6.7\times$\textasciitilde$41.7\times$ and $1.9\times$\textasciitilde$11.9\times$ higher than OuterSPACE, respectively. \\
	\indent\setlength{\parindent}{1em}For GCNAX, it builds a specilized accelerator for SDMM, which can remove a part of operations in SSMM and consumes less power. However, the $1\times16$ PE array of GCNAX consumes higher bandwidth on average compared with the $8\times8$ PE array of IOPS, inducing more storage resources. Thus, we achieve $3.3\times$ $R_T$ compared with GCNAX but our $E_{T}$ is $0.4\times$ lower then it. Finally, our $E_N$ and $R_N$ are $0.3\times$\textasciitilde$4.9\times$ and $2.4\times$\textasciitilde$40\times$ higher than GCNAX, respectively. \\
	\indent\setlength{\parindent}{1em}For OuterSPACE, its PE structure is more general-purpose, containing many caches and instruction decoding modules, while our architecture is specialized for SpMM with simpler computing flow. Therefore, our $E_T$ and $R_T$ are $2.7\times$ and $4.4\times$ higher than OuterSPACE, respectively. Finally, our energy and resource efficiency are $4.5\times$\textasciitilde$10.5\times$ and $8.3\times$\textasciitilde$19.4\times$ higher than OuterSPACE, respectively. \\
	\indent\setlength{\parindent}{1em}For SpArch, most of power and area are consumed by the merge tree, which is similar with the PE array in IOPS. Both of us aim to relieve the irregularity of psum access. Therefore, our $E_T$ and $R_T$ are $1.3\times$ and $0.9\times$ higher than SpArch, respectively. Combining the performance, our $E_N$ and $R_N$ are $1.7\times$\textasciitilde$6.3\times$ and $1.2\times$\textasciitilde$4.4\times$ higher compared with SpArch, respectively.

	\subsection{DRAM Access}
\begin{figure}[H]
		\centering
		\includegraphics[width=0.99\linewidth]{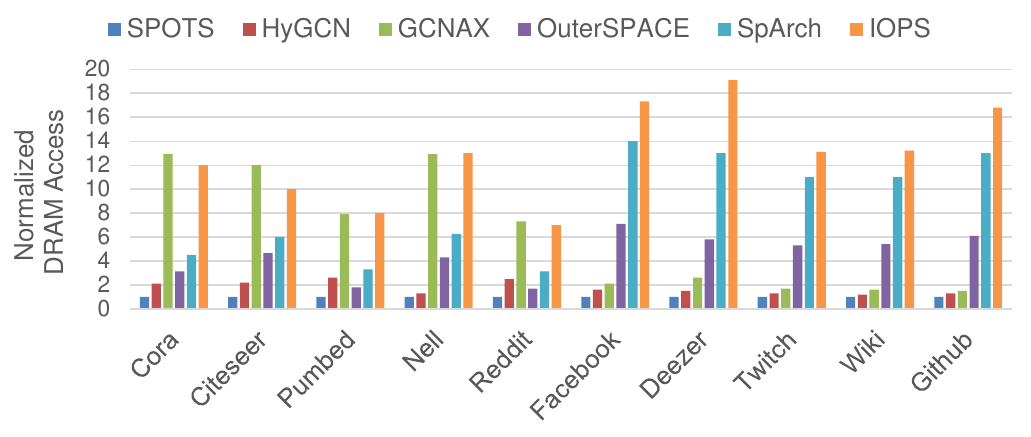}
		\caption{DRAM access comparison with SPOTS, OuterSPACE, HyGCN, GCNAX, and SpArch.}
		\label{fig14}
	\end{figure}

	\indent\setlength{\parindent}{1em}To explore the energy consumption of off-chip data movement, we compare DRAM access saving with SPOTS, HyGCN, GCNAX, OuterSPACE, and SpArch, achieving $7\times$\textasciitilde$19.1\times$, $2.8\times$\textasciitilde$12.9\times$, $0.8\times$\textasciitilde$11.2\times$, $2.2\times$\textasciitilde$4.4\times$, and $1.2\times$\textasciitilde$2.7\times$, respectively, as shown in Fig.\ref{fig14}. To make a fair comparison, we assume the same on-chip storage to explore the difference of dataflow in each accelerator. The specific parameters of matrix partition are shown in Tab.\ref{tab10}, whose meanings have been explained in Tab.\ref{tab5}. For SDMM benchmark, we show the partition of two steps in the first layer, where we reuse the sparse matrices $A$ first (RAF) in most case, except for the Pumbed and Reddit dataset with the dense feature matrix. For the SSMM benchmark, we reuse each matrix based on the DRAM access in each case. The average utilization of input buffer is $50\%$\textasciitilde$75\%$, fully utilizing the buffers without overflowing. Since the psum matrices in SDMM are dense, we can achieve a 100\% utilization theoretically.  \\
	\indent\setlength{\parindent}{1em}For SPOTS, though there are many input and output matrices reuse in the PE array, the mismatched index of input matrices induces large amount of zeros and wastes the buffer storage to store the zero element. The more zeros contained in the buffer, the smaller sizes of submatrix are, leading to more tiling blocks and repeat DRAM access of input matrices. Thus, we achieve $7\times$\textasciitilde$19.1\times$ DRAM access saving. \\
	\indent\setlength{\parindent}{1em}For HyGCN, the SSMM engine stations the input matrix A and shrinks the submatrix size to reduce the sparsity. Then, it repeatedly  loads the corresponding element of input matrix B. However, for the SDMM benchmark, the execution order brings more computation with more data access from DRAM. For the SSMM benchmark, the output matrix is stored in a dense format as the input for DDMM engine, causing the storage waste and lowering the data reusing rate. Therefore, we reduce $2.8\times$\textasciitilde$12.9\times$ DRAM access. \\
	\indent\setlength{\parindent}{1em}For GCNAX, it reuses the sparse matrix and repeatly accesses the dense matrix in the SDMM computing, similar to our approach in this case. However, for the SSMM benchmark, the accumulated psum matrix can contain many zero elements, which limits the sizes of output submatrices. The DRAM access can be more frequent due to more tiling blocks. Thus, we achieve $0.8\times$\textasciitilde$11.2\times$ DRAM access saving. \\
	\indent\setlength{\parindent}{1em}For OuterSPACE, the main DRAM access comes from the irregular psum accumulation. Compared with OuterSPACE, we relieve the irregularity of the psum access by applying address mapping, which obtains the final output matrix on-chip without repeat DRAM access. Therefore, we achieve $2.2\times$\textasciitilde$4.4\times$ DRAM access saving. \\
	\indent\setlength{\parindent}{1em}For SpArch, it reduces the psum matrix by a Huffman tree scheduler and merges the psum to calculate final output matrix, greatly reducing the DRAM access of psum accumulation. However, there is no input reuse between each PE, which causes more repeat access of input matrix and gets worse in the SDMM benchmark. Besides, we reorganize the psum buffer for $2\times$ storage in SDMM, reducing DRAM access further. Thus, we reduce $1.2\times$\textasciitilde$2.7\times$ DRAM access. 
	
	\begin{table*}\centering
		\caption{Parameters of Matrix Partition}
		\label{tab10}
		\begin{tabular}{|c|c|c|c|c|c|c|c|c|c|}
			\hline
			Dataset  & Mt      & Nt  & Kt          & Tk    & Tm      & Tn     & \begin{tabular}[c]{@{}c@{}}Buffer Utilization\\ (A/B/Psum)\end{tabular} & \begin{tabular}[c]{@{}c@{}}Computing \\ Mode\end{tabular} & \begin{tabular}[c]{@{}c@{}}Reuse\\ Strategy\end{tabular} \\ \hline
			cora     & 339,339 & 1,1 & 2708,287    & 1,5   & 1,1     & 180,2  & 63\%/62\%/66\%                                                          & SDMM,SDMM                                                 & RAF,RAF                                                  \\ \hline
			citeseer & 416,416 & 1,1 & 3327,412    & 1,9   & 1,1     & 463,2  & 56\%/71\%/81\%                                                          & SDMM,SDMM                                                 & RAF,RAF                                                  \\ \hline
			pumbed   & 493,4   & 1,2 & 9859,500    & 2,1   & 5,617   & 63,1   & 55\%/75\%/96\%                                                          & SDMM, DDMM                                                & RABE,RBF                                                 \\ \hline
			nell     & 504,504 & 1,1 & 42336,20426 & 4,3   & 42,42   & 7660,8 & 65\%/55\%/98\%                                                          & SDMM, SDMM                                                & RAF,RAF                                                  \\ \hline
			Reddit   & 511,10  & 1,8 & 1421,201    & 164,3 & 57,2912 & 76,1   & 74\%/74\%/91\%                                                          & SDMM, DDMM                                                & RABE,RAF                                                 \\ \hline
			Facebook & 201     & 43  & 11235       & 2     & 14      & 14     & 75\%/71\%/74\%                                                          & SSMM                                                      & RAF                                                      \\ \hline
			Deezer   & 442     & 83  & 14141       & 2     & 4       & 47     & 67\%/63\%/67\%                                                          & SSMM                                                      & RAF                                                      \\ \hline
			Twitch   & 92      & 14  & 9498        & 1     & 13      & 24     & 72\%/53\%/75\%                                                          & SSMM                                                      & RBF                                                      \\ \hline
			Wiki     & 112     & 23  & 5816        & 2     & 13      & 72     & 41\%/74\%/70\%                                                          & SSMM                                                      & RBF                                                      \\ \hline
			Github   & 197     & 27  & 9425        & 4     & 24      & 19     & 73\%/57\%/71\%                                                          & SSMM                                                      & RAF                                                      \\ \hline
		\end{tabular}
	\end{table*}
	
	\section{Design Space Exploration}
\begin{figure}[H]
		\centering
		\includegraphics[width=0.99\linewidth]{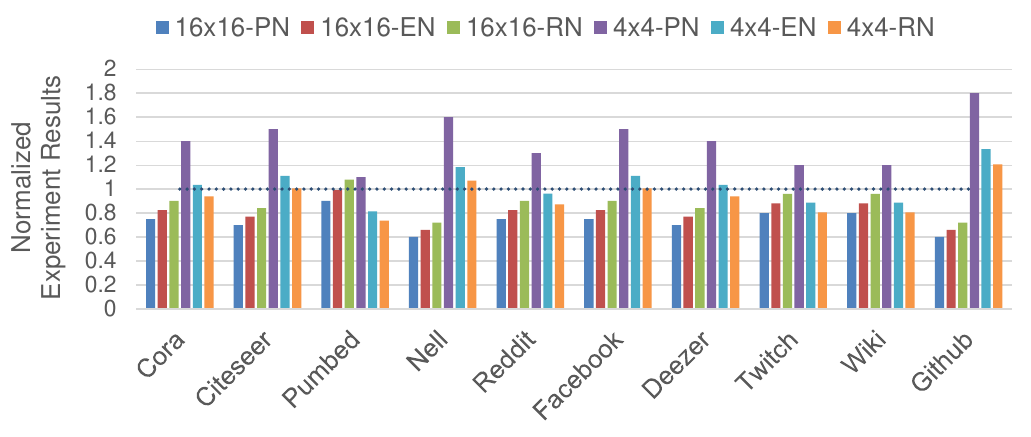}
		\caption{Experiment results comparison among different PE array sizes.}
		\label{fig15}
	\end{figure}

\begin{figure}[H]
		\centering
		\includegraphics[width=0.99\linewidth]{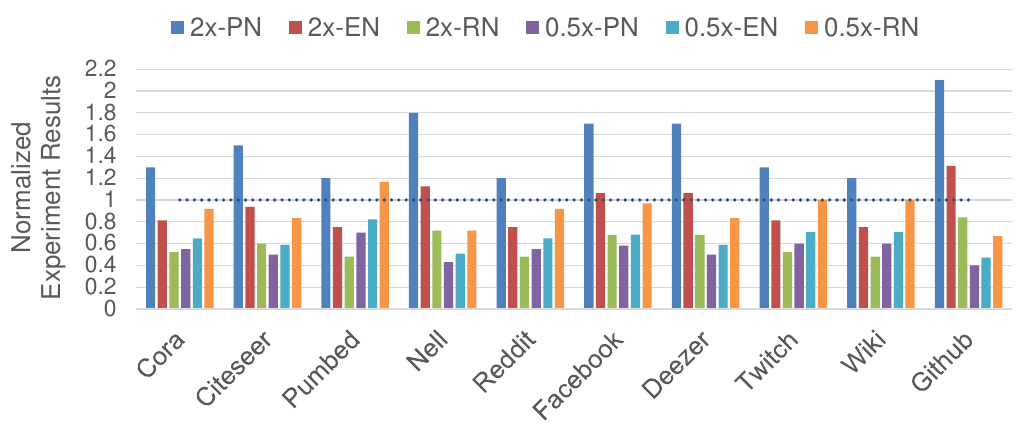}
		\caption{Experiment results comparison among different buffer storages.}
		\label{fig16}
	\end{figure}

	To explore the influence of hardware parameters on performance, energy efficiency, and resource efficiency, we testify the IOPS with different PE array size and buffer storage, as shown in Fig.\ref{fig15} and Fig.\ref{fig16}. Since the DRAM access is mainly determined by the dataflow, there is no need to explore in the same architecture. For the PE array size, $G_{NA}$x$G_{NB}$,  the area of PE array and input buffer in the architecture are scaled by square and linearity, respectively, indicating that the larger $G_{NA}$x$G_{NB}$ is, the higher $E_{T}$ and $R_{T}$ we achieve. However, with the increasing of PE array size, the density of each submatrix in each PE can be more imbalanced, causing performance degradation. Thus, we set three different $G_{NA}$x$G_{NB}$, $4\times4$, $8\times8$, and $16\times16$, to explore their difference. As shown in Fig.\ref{fig15}, the $8\times8$ PE array achieves $0.56\times$\textasciitilde$ 0.92\times$ performance with $1.35\times$ $E_{T}$ and $1.5\times$ $R_{T}$ compared with $4\times4$ PE array, and achieves $1.15\times$\textasciitilde$ 1.68\times$ performance with $0.91\times$ $E_{T}$ and $0.83\times$ $R_{T}$ compared with $16\times16$ PE array. Considering the final $E_N$ and $R_N$, $8\times8$ PE array performs better in these benchmarks.  \\
	\indent\setlength{\parindent}{1em}For the buffer storage, the larger submatrix we can store, the more balanced density of each submatrix is, indicating higher performance. However, with larger buffer comes more resource and power overhead. Thus, we shrink and magnify the input buffer $2\times$ (psum buffer $4\times$) to testify the benchmark, where the $E_T$ of $2\times$ buffer and $0.5\times$ buffer are $0.63\times$ and $1.2\times$ that of the basic buffer, respectively, and the $R_T$ are $0.4\times$ and $1.7\times$, respectively. Combining with the $P_N$, the $E_N$ of $1\times$ buffer is $1.3\times$\textasciitilde$ 2.1\times$ and $0.76\times$\textasciitilde$ 1.3\times$ higher than $0.5\times$ buffer and $2\times$ buffer, respectively, and the $R_N$ is $0.86\times$\textasciitilde$ 1.5\times$ and $1.2\times$\textasciitilde$ 2.1\times$, respectively. Therefore, we choose $1\times$ buffer for the balance of performance and hardware overhead.
	
	\section{Conclusion}
	This paper proposes an IOHP based accelerator for SpMM, which reuses the input matrix across PEs with inner product based dataflow, and removes zero-element calculations based on outer product. Besides, an address mapping method is designed to address the irregularity of DRAM access during psum accumulation. Furthermore, an adaptive partition strategy is proposed to tile the input matrices based on their sparsity ratio, effectively utilizing the storage of architecture and reducing DRAM access. Collaborating with the above method, this accelerator can switch the computing mode between the SSMM and SDMM efficiently. Compared with SpArch, we achieve $1.7\times$\textasciitilde$6.3\times$ energy efficiency and $1.2\times$\textasciitilde$4.4\times$ resource efficiency, with $1.4\times$\textasciitilde$2.1\times$ DRAM access saving. \\

	%
	
	\bibliographystyle{IEEEtran}
	\bibliography{saprse_matrix}

\begin{thebibliography}{10}
\providecommand{\url}[1]{#1}
\csname url@samestyle\endcsname
\providecommand{\newblock}{\relax}
\providecommand{\bibinfo}[2]{#2}
\providecommand{\BIBentrySTDinterwordspacing}{\spaceskip=0pt\relax}
\providecommand{\BIBentryALTinterwordstretchfactor}{4}
\providecommand{\BIBentryALTinterwordspacing}{\spaceskip=\fontdimen2\font plus
\BIBentryALTinterwordstretchfactor\fontdimen3\font minus
  \fontdimen4\font\relax}
\providecommand{\BIBforeignlanguage}[2]{{%
\expandafter\ifx\csname l@#1\endcsname\relax
\typeout{** WARNING: IEEEtran.bst: No hyphenation pattern has been}%
\typeout{** loaded for the language `#1'. Using the pattern for}%
\typeout{** the default language instead.}%
\else
\language=\csname l@#1\endcsname
\fi
#2}}
\providecommand{\BIBdecl}{\relax}
\BIBdecl

\bibitem{10.5555/3118786.3119254}
P.~D'alberto and A.~Nicolau, ``R-kleene: A high-performance divide-and-conquer
  algorithm for the all-pair shortest path for densely connected networks,''
  \emph{Algorithmica}, vol.~47, no.~2, p. 203–213, feb 2007.

\bibitem{10.1145/3219819.3219890}
R.~Ying, R.~He, and et~al, ``Graph convolutional neural networks for web-scale
  recommender systems,'' in \emph{ACM KDD}, 2018, p. 974–983.

\bibitem{10.5555/866839}
S.~Goedecker and L.~Colombo, ``Tight binding molecular dynamics on parallel
  computers,'' USA, Tech. Rep., 1994.

\bibitem{kipf2017semisupervised}
\BIBentryALTinterwordspacing
T.~N. Kipf and M.~Welling, ``Semi-supervised classification with graph
  convolutional networks,'' in \emph{arXiv}, 2017. [Online]. Available:
  \url{https://arxiv.org/abs/1609.02907}
\BIBentrySTDinterwordspacing

\bibitem{snapnets}
J.~Leskovec and A.~Krevl, ``{SNAP Datasets}: {Stanford} large network dataset
  collection,'' \url{http://snap.stanford.edu/data}, Jun. 2014.

\bibitem{8192463}
N.~P. Jouppi, C.~Young, and et~al, ``In-datacenter performance analysis of a
  tensor processing unit,'' in \emph{ISCA}, 2017, pp. 1--12.

\bibitem{9586257}
G.~Jeong, E.~Qin, and et~al, ``Rasa: Efficient register-aware systolic array
  matrix engine for cpu,'' in \emph{DAC}, 2021, pp. 253--258.

\bibitem{94}
Y.~Zhu and et~al, ``Exploiting parallelism with vertex-clustering in
  processing-in-memory-based gcn accelerators,'' in \emph{DATE}, 2022, pp.
  652--657.

\bibitem{9774537}
------, ``Exploiting parallelism with vertex-clustering in
  processing-in-memory-based gcn accelerators,'' in \emph{DATE}, 2022, pp.
  652--657.

\bibitem{9065592}
M.~Yan, L.~Deng, X.~Hu, L.~Liang, Y.~Feng, X.~Ye, Z.~Zhang, D.~Fan, and Y.~Xie,
  ``Hygcn: A gcn accelerator with hybrid architecture,'' in \emph{2020 IEEE
  International Symposium on High Performance Computer Architecture (HPCA)},
  2020, pp. 15--29.

\bibitem{10.1145/3532863}
M.~Soltaniyeh and et~al, ``An accelerator for sparse convolutional neural
  networks leveraging systolic general matrix-matrix multiplication,''
  \emph{ACM Trans. Archit. Code Optim.}, vol.~19, no.~3, may 2022.

\bibitem{49}
\BIBentryALTinterwordspacing
S.~Dalton, L.~Olson, and N.~Bell, ``Optimizing sparse matrix—matrix
  multiplication for the gpu,'' \emph{ACM Trans. Math. Softw.}, vol.~41, no.~4,
  oct 2015. [Online]. Available: \url{https://doi.org/10.1145/2699470}
\BIBentrySTDinterwordspacing

\bibitem{51}
W.~Liu and B.~Vinter, ``An efficient gpu general sparse matrix-matrix
  multiplication for irregular data,'' in \emph{2014 IEEE 28th International
  Parallel and Distributed Processing Symposium}, 2014, pp. 370--381.

\bibitem{lee}
J.~Lee, Kang, and et~al, ``Optimization of gpu-based sparse matrix
  multiplication for large sparse networks,'' in \emph{ICDE}, 2020, pp.
  925--936.

\bibitem{10089919}
N.~G, N.~S, and K.~S, ``Moscon: Modified outer product based sparse
  matrix-matrix multiplication accelerator with configurable tiles,'' in
  \emph{VLSID}, 2023, pp. 264--269.

\bibitem{8327050}
S.~Pal, J.~Beaumont, and et~al, ``Outerspace: An outer product based sparse
  matrix multiplication accelerator,'' in \emph{HPCA}, 2018, pp. 724--736.

\bibitem{9065428}
Z.~Zhang, H.~Wang, and et~al, ``Sparch: Efficient architecture for sparse
  matrix multiplication,'' in \emph{HPCA}, 2020, pp. 261--274.

\bibitem{9252000}
T.~Geng, A.~Li, R.~Shi, C.~Wu, T.~Wang, Y.~Li, P.~Haghi, A.~Tumeo, S.~Che,
  S.~Reinhardt, and M.~C. Herbordt, ``Awb-gcn: A graph convolutional network
  accelerator with runtime workload rebalancing,'' in \emph{2020 53rd Annual
  IEEE/ACM International Symposium on Microarchitecture (MICRO)}, 2020, pp.
  922--936.

\bibitem{9407104}
J.~Li, A.~Louri, A.~Karanth, and R.~Bunescu, ``Gcnax: A flexible and
  energy-efficient accelerator for graph convolutional neural networks,'' in
  \emph{2021 IEEE International Symposium on High-Performance Computer
  Architecture (HPCA)}, 2021, pp. 775--788.

\bibitem{9251978}
N.~Srivastava, H.~Jin, and et~al, ``Matraptor: A sparse-sparse matrix
  multiplication accelerator based on row-wise product,'' in \emph{MICRO},
  2020, pp. 766--780.

\bibitem{10.1145/1137856.1137866}
\BIBentryALTinterwordspacing
H.~Kaplan, M.~Sharir, and E.~Verbin, ``Colored intersection searching via
  sparse rectangular matrix multiplication,'' in \emph{Proceedings of the
  Twenty-Second Annual Symposium on Computational Geometry}, ser. SCG
  '06.\hskip 1em plus 0.5em minus 0.4em\relax New York, NY, USA: Association
  for Computing Machinery, 2006, p. 52–60. [Online]. Available:
  \url{https://doi.org/10.1145/1137856.1137866}
\BIBentrySTDinterwordspacing

\bibitem{10.1109/MCSE.2008.45}
\BIBentryALTinterwordspacing
J.~R. Gilbert, S.~Reinhardt, and V.~B. Shah, ``A unified framework for
  numerical and combinatorial computing,'' \emph{Computing in Science and
  Engg.}, vol.~10, no.~2, p. 20–25, mar 2008. [Online]. Available:
  \url{https://doi.org/10.1109/MCSE.2008.45}
\BIBentrySTDinterwordspacing

\bibitem{RABIN1989557}
\BIBentryALTinterwordspacing
M.~O. Rabin and V.~V. Vazirani, ``Maximum matchings in general graphs through
  randomization,'' \emph{Journal of Algorithms}, vol.~10, no.~4, pp. 557--567,
  1989. [Online]. Available:
  \url{https://www.sciencedirect.com/science/article/pii/0196677489900059}
\BIBentrySTDinterwordspacing

\bibitem{10.5555/3294771.3294869}
W.~L. Hamilton, R.~Ying, and J.~Leskovec, ``Inductive representation learning
  on large graphs,'' in \emph{Proceedings of the 31st International Conference
  on Neural Information Processing Systems}, ser. NIPS'17.\hskip 1em plus 0.5em
  minus 0.4em\relax Red Hook, NY, USA: Curran Associates Inc., 2017, p.
  1025–1035.

\bibitem{Xu2018HowPA}
\BIBentryALTinterwordspacing
K.~Xu, W.~Hu, J.~Leskovec, and S.~Jegelka, ``How powerful are graph neural
  networks?'' \emph{ArXiv}, vol. abs/1810.00826, 2018. [Online]. Available:
  \url{https://arxiv.org/abs/1810.00826}
\BIBentrySTDinterwordspacing

\bibitem{GAT}
\BIBentryALTinterwordspacing
P.~Veličković, G.~Cucurull, A.~Casanova, A.~Romero, P.~Liò, and Y.~Bengio,
  ``Graph attention networks,'' 2018. [Online]. Available:
  \url{https://arxiv.org/abs/1710.10903}
\BIBentrySTDinterwordspacing

\bibitem{Cora}
A.~McCallum, ``Cora,'' \url{http://www.cs.umd.edu/~sen/lbc-proj/LBC.html}.

\bibitem{10.1145/276675.276685}
C.~L. Giles, K.~D. Bollacker, and S.~Lawrence, ``Citeseer: An automatic
  citation indexing system,'' in \emph{Proceedings of the Third ACM Conference
  on Digital Libraries}, ser. DL '98, 1998, p. 89–98.

\bibitem{Sen2008CollectiveCI}
P.~Sen, G.~Namata, and et~al, ``Collective classification in network data,'' in
  \emph{The AI Magazine}, 2008.

\bibitem{10.5555/2898607.2898816}
A.~Carlson, J.~Betteridge, and et~al, ``Toward an architecture for never-ending
  language learning,'' in \emph{Proceedings of the Twenty-Fourth AAAI
  Conference on Artificial Intelligence}.\hskip 1em plus 0.5em minus
  0.4em\relax AAAI Press, 2010, p. 1306–1313.

\bibitem{10.1145/2783258.2783417}
\BIBentryALTinterwordspacing
P.~Yanardag and S.~Vishwanathan, ``Deep graph kernels,'' in \emph{Proceedings
  of the 21th ACM SIGKDD International Conference on Knowledge Discovery and
  Data Mining}, ser. KDD '15.\hskip 1em plus 0.5em minus 0.4em\relax New York,
  NY, USA: Association for Computing Machinery, 2015, p. 1365–1374. [Online].
  Available: \url{https://doi.org/10.1145/2783258.2783417}
\BIBentrySTDinterwordspacing

\bibitem{7013055}
O.~Villa, D.~R. Johnson, M.~Oconnor, E.~Bolotin, D.~Nellans, J.~Luitjens,
  N.~Sakharnykh, P.~Wang, P.~Micikevicius, A.~Scudiero, S.~W. Keckler, and
  W.~J. Dally, ``Scaling the power wall: A path to exascale,'' in \emph{SC '14:
  Proceedings of the International Conference for High Performance Computing,
  Networking, Storage and Analysis}, 2014, pp. 830--841.

\end{thebibliography}

\end{document}